\documentclass[twocolumn,brazil,english,aps,eqsecnum,showpacs,prb]{revtex4}

\usepackage{amssymb}
\usepackage{pslatex}
\usepackage[T1]{fontenc}
\usepackage[latin1]{inputenc}
\usepackage{graphicx}
\usepackage{graphicx,epsf}
\usepackage{babel}
\usepackage{amsmath}



\begin{document}

\title{Forward Scattering Approximation and Bosonization in Integer Quantum Hall
Systems}
\author{M. Rosenau da Costa,$^{1,2}$ H. Westfahl Jr.,$^{3}$ and A.O. Caldeira$^{2}$}
\date{\today{}}

\affiliation{$^{1}$International Center of Condensed Matter Physics, Universidade de Bras\'ilia, Caixa Postal 04667, 70910-900, Bras\'ilia-DF, Brazil\\
$^{2}$Instituto de F\'isica {}``Gleb Wataghin'', Universidade
Estadual de Campinas,
Caixa Postal 6165, 13083-970, Campinas-SP, Brazil\\
 $^{3}$Laborat{\'o}rio Nacional de Luz S{\'\i}ncrotron - ABTLuS, Caixa Postal 6192, 13043-090, Campinas-SP, Brazil}

\begin{abstract}
In this work we present a model and a method to study
integer quantum Hall (IQH) systems. Making use of the Landau
levels structure we divide these two dimensional
systems into a set of interacting one dimensional gases, one
for each guiding center. We show that the
so-called \textit{strong field approximation}, used by Kallin
and Halperin and by MacDonald, is equivalent, in first order, to a \textit{forward scattering approximation} and analyze the IQH systems
within this approximation. Using an appropriate variation of the
Landau level bosonization method we obtain the dispersion relations for the collective excitations and the single particle spectral functions. For the bulk states, these results evidence a behavior
typical of \textit{non-normal} strongly correlated systems,
including the spin-charge splitting of the single particle
spectral function. We discuss the origin of this behavior in the
light of the Tomonaga-Luttinger model and the bosonization of two
dimensional electron gases.
\end{abstract}

\maketitle

\section{Introduction}

The integer quantum Hall effect (IQHE) \cite{Prange} appears in
two dimensional electron gases (2DEG) under a strong magnetic
field, $B$, with the electronic Landau level filling factor,
$\nu$, equals to an integer number. In this case the
non-interacting ground state presents $\nu$
 completely filled  Landau levels and the excitons generated by the
electron-electron interaction involve particles in different
levels. At stronger magnetic fields the degeneracy,
$N_{\phi}=S/2\pi l^{2}$ (where $S$ is the area of the 2DEG and
$l=\sqrt{c\hbar/(eB)}$ is the magnetic length), of a single Landau
level exceeds the number of electrons. In this case many
non-interacting ground states can be constructed with all the
electrons in the macroscopically degenerate lowest Landau level.
In this case the usual perturbative theories, which lead to the
Fermi Liquid behavior, are invalid and the effect of the
electron-electron interaction turns out to be much more dramatic,
giving rise to a new state of the matter and to the fractional
quantum Hall effect (FQHE) \cite{Laug0,Sarma}.

The study of collective excitations plays an important role in
understanding both systems. The
existence of magneto-rotons \cite{Halp,Kallin,MacDH} was verified through
 inelastic light scattering \cite{PinkKal} and phonon spectroscopy \cite{Zeit,Apalk}. However, in the integer quantum Hall systems, at very
low electronic densities (less than $10^{-10}cm^{-2}$) multiple
magneto-rotons which are not accessible to the usual perturbative
theories (possibly because they neglect the mixing of the Landau
levels) were observed \cite {Eric}, demonstrating an incomplete
understanding of these excitations. Additionally, in the FQHE the
coupling between the particle
excitations with the magneto-plasmon mode close to the cyclotron frequency, $%
\omega _{c}$, seems to be the key for a Hamiltonian theory of these systems
\cite{ShaN}, and to the origin of most part of the
notable properties of their quasi-particles, the \textit{composite fermions}
\cite{Jain,Hei}. Here we will show that the collective excitations also seem
to have a strong influence in the bulk quasi-particles properties even in the
IQHE.

In this work we will explicitly treat the integer quantum Hall
systems, although the method we present, together with the
Chern-Simons theory \cite {Read}, also has applications in the
fractional case \cite{Mar}. Assuming a system where the Coulomb
energy, $e^{2}/\varepsilon l$, is much smaller than the cyclotron
energy, $\hbar \omega _{c}=\hbar eB/(mc)$, the so-called
\textit{strong field approximation} \cite{Halp,MacDH} (where the
mixing between the Landau levels is neglected) can be used. Under
this approximation, the collective magneto-plasma modes were
determined by the Green's function method \cite{Halp} and
time-dependent Hartree-Fock
 theory \cite{MacDH}. In this work, we will see that these
modes can also be obtained by a magneto-exciton bosonization
method. First, making use of the natural structure of the Landau
levels, we divide the 2DEG into a set of $N_{\phi }$ one
dimensional interacting channels, one for each guiding center occupied by $%
\nu $ particles. We show that, to first order in $\left(
e^{2}/\varepsilon l\right) /$ $\hbar \omega _{c}$, the \textit{strong field
approximation} is equivalent to a \textit{forward scattering approximation}
in this model of one dimensional interacting channels. Thus, we will allow
for the transfer of energy and momentum between the
different channels mediated by the direct and exchange Coulomb interactions, conserving
the net number of particles in each channel. We then use a variation of the \textit{Landau level bosonization%
} \cite{Harry} which is ideal for a treatment of the interaction
effects in this model.

One of the major advantages of this bosonization method is that,
through the \textit{Mattis-Mandelstam} bosonic representation of
the fermionic operator \cite{mandelstam,VanD}, one is be able to
determine the single particle properties directly from the
collective modes, differently from the
usual perturbative theory. Another advantage is that this \textit{%
Mattis-Mandelstam} representation is an identity \cite{VanD} which, technicaly,
can be used later to extend the model, including
the net transfer of particles between neighbor channels according to their
relative angular momentum. We believe that bosonization procedures
similar to this one can be important tools to clarify the relation
between the quasi-particle properties and the collective modes in
the FQHE. Indeed, the Hamiltonian Theory of the FQHE \cite{ShaN}
is very similar to a bosonization of the degrees of freedom
related to the magneto-plasmon excitation.

The organization of the paper is as follows. In section II we review the
free and the interacting Hamiltonians of a two dimensional electron gas in a
perpendicular magnetic field written in the Landau level basis and in the
symmetric gauge. In section III we show the equivalence between the \textit{%
strong field approximation} and the \textit{forward scattering
approximation} and then describe how to treat the 2DEG, including
the direct and exchange interactions between the electrons in a
bosonization approach. In section IV we present a comparative
analysis of our bosonized model with the
\textit{Tomonaga-Luttinger} model and the 2D bosonization. Next we
describe the strategy to deal with spin-full fermions in section
V. The spectrum of neutral excitations  obtained is presented in
section VI. The construction of the fermionic operator and the
determination of the single-particle spectral functions is
accomplished in section VII and our conclusions are presented in
section VIII.

\section{Fermionic Representation}

We consider a 2DEG in a strong perpendicular magnetic field at
zero temperature with an integer number, $\nu$, of fully occupied
Landau levels. We start from a simpler description of spinless
electrons and generalize the model to the case of electrons with
spins in section VII.

The non-interacting part of the Hamiltonian of spinless fermions of mass $m$
in a plane perpendicular to an uniform magnetic field is given by
\begin{equation}
H_{0}=\frac{1}{2m}\int d^{2}\mathbf{r}\Psi ^{\dagger }(\mathbf{r})\left[
\mathbf{p}-\frac{e}{c}\mathbf{A}(\mathbf{r})\right] ^{2}\Psi (\mathbf{r}),
\label{H0cam}
\end{equation}
where $\Psi \left( \mathbf{r}\right) $ is the fermion field operator, which
can be expanded in the Landau level basis and in the symmetric gauge, $%
\mathbf{A}\left( \mathbf{r}\right) =\frac{B}{2}\hat{z}\times \mathbf{r}$, as
\begin{equation}
\Psi (\mathbf{r})=\sum_{n,m}\langle \mathbf{r}|n,m\rangle c_{n,m}.
\label{expferm}
\end{equation}
Here $|n,m\rangle $ are the eigenstates of the one particle Hamiltonian,
where $n$ refers to the Landau level and $m$ to the guiding center, with
\cite{allan}
\begin{equation}
\langle \mathbf{r}|n,m\rangle =\frac{e^{-z^{2}/4l^{2}}}{\sqrt{2\pi l^{2}}}%
G_{m,n}\left( i\frac{\mathbf{z}^{\ast }}{l}\right) ,  \label{G}
\end{equation}
where $\mathbf{z}=x+iy$. For $m>n$,
\begin{equation}
G_{m,n}(\mathbf{z})=\sqrt{\frac{n!}{m!}}\left( \frac{-i\mathbf{z}}{\sqrt{2}}%
\right) ^{m-n}L_{n}^{m-n}\left( \frac{z^{2}}{2}\right) ,  \label{Gdef}
\end{equation}
with $L_{n}^{m-n}$ being the generalized Laguerre polynomial. The
Hamiltonian $H_{0}$ is diagonal in this Landau levels basis
\begin{equation}
H_{0}=\hbar \omega _{c}\sum_{m=0}^{N_{\Phi }-1}\sum_{n=0}^{\infty
}(n+1/2)c_{n,m}^{\dagger }c_{n,m}.  \label{H0}
\end{equation}
Given the guiding center degeneracy, $N_{\phi }$, the non-interacting ground
state is defined by uniformly filling $\nu $ Landau levels of each guiding
center $m:$%
\begin{equation*}
\left| G_{\mathbf{0}}\right\rangle =\prod_{m=0}^{N_{\phi
}-1}\prod_{n=0}^{\nu -1}c_{n,m}^{\dagger }\left| 0\right\rangle ,
\end{equation*}
where $N_{\phi }=S/2\pi l^{2}$ is the number of flux quanta crossing the
area of the system, $S$. The neutral excitations in the non-interacting
system are electron-hole pairs with a hole in a Landau level $p\leq \nu -1$
at the guiding center $m$ and an electron in a level $p+n>\nu -1$ at the
guiding center $m^{\prime }$.

In the Landau levels basis the Coulomb interaction, $V\left( r\right)
=e^{2}/\varepsilon r$, is formally given by
\begin{eqnarray}
H_{I} &=&\frac{1}{2}\sum_{\left\{ n,m\right\} }\langle
n_{1},m_{1};n_{2},m_{2}|V|n_{1}^{\prime },m_{1}^{\prime };n_{2}^{\prime
},m_{2}^{\prime }\rangle  \notag \\
&&\times c_{n_{1},m_{1}}^{\dagger }c_{n_{2},m_{2}}^{\dagger
}c_{n_{2}^{\prime },m_{2}^{\prime }}c_{n_{1}^{\prime },m_{1}^{\prime }},
\label{HI}
\end{eqnarray}
where the interaction matrix element can be written in the momentum
representation as
\begin{eqnarray*}
\langle n_{1},m_{1};n_{2},m_{2}|V|n_{1}^{\prime },m_{1}^{\prime
};n_{2}^{\prime },m_{2}^{\prime }\rangle &=&\frac{1}{\left( 2\pi \right) ^{2}%
}\int d^{2}q\tilde{V}\left( q\right) e^{-l^{2}q^{2}} \\
&&\times G_{n_{1}^{\prime },n_{1}}^{\ast }\left( l\mathbf{q}^{\ast }\right)
G_{m_{1}^{\prime },m_{1}}^{\ast }\left( l\mathbf{q}\right) \\
&&\times G_{n_{2},n_{2}^{\prime }}\left( l\mathbf{q}^{\ast }\right)
G_{m_{2},m_{2}^{\prime }}\left( l\mathbf{q}\right) ,
\end{eqnarray*}
with $\tilde{V}\left( q\right) =2\pi e^{2}/\varepsilon q$ and $G$ defined in
(\ref{Gdef}).

\section{The Forward Scattering Approximation and the Bosonic Description}

\subsection{First Order Equivalence Between the Strong Field Approximation
and the Forward Scattering Approximation}

\begin{figure}[tbp]
 \includegraphics[width=1.0%
\columnwidth,keepaspectratio]{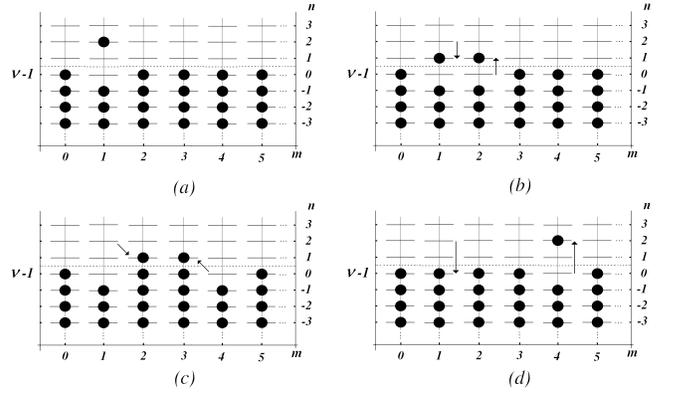}
\par
 \caption{(a) A single exciton created over a ground state that completely fills the $%
\nu -1$ lowest Landau levels, neglecting any Landau level mixing (ii).
Consider now only first order processes, where the total energy is
conserved, $\sum_{i}\Delta n_{i}=0$. If we neglect multiexciton states (i),
processes like (b) or (c), which conserve the total angular momentum in the
symmetric gauge, $\Delta L_{z}=\sum_{i}\left( \Delta n_{i}-\Delta
m_{i}\right) =0$, will not be allowed. The only dynamical process allowed
for this exciton (in first order and constrained to (i) and (ii)) is to
decay in its own guiding center and create another exciton in a different
one, as in (d). So, the number of particles in each guiding center $m$ will
be kept constant. The dispersion curves that we will obtain for the excitons
are related to the breakdown of degeneracy between states like (a) and (d).
Although the forward scattering approximation allows the decay processes
like (b), we will not consider then due to an additional approximation made
in the bosonization method.} \label{Fig1}
\end{figure}

Inspired by the linear energy dispersion of the quantum number $n$
and the degeneracy of the quantum number $m$, one can make an
analogy between the quantum Hall system and the problem of many
interacting one-dimensional electron gases. In fact, the
elementary neutral excitations spectrum in the integer quantum
Hall regime, consisting of electron-hole pairs excited over a
completely filled set of Landau levels, has been calculated using
the diagrammatic Green's function approach of Kallin and Halperin
\cite{Halp} and the time-dependent Hartree-Fock approximation of
MacDonald \cite{MacDH}. In both methods the excited states are
constructed essentially in a single mode approximation: (i)
considering \ only the presence of a single electron-hole pair at
a time, (ii) over a ground state that neglects any Landau level
mixing. Under these approximations, named {\it strong field
approximation}, the magnetoplasmon bands were calculated including
different order terms in the interaction strength, $\gamma =\left(
e^{2}/\varepsilon l\right) /$ $\hbar \omega _{c}\ll 1$. Kallin and
Halperin include only some first order terms, not all of them,
once they do not consider the ones associated with the decay of
one exciton into two. MacDonald includes some second order terms,
although also excluding the first order ones neglected by Kallin
and Halperin. The effect of the decay of excitons was partially
considered by Cheng \cite{Cheng}. In Fig.(\ref{Fig1}), which was
constructed according to the conditions (i)\ and (ii)\ above, we
show that, to first order in $\gamma$, the strong field
approximation is analogous to the {\it forward scattering
approximation} in a set of one-dimensional electron gases, where
the net number of particles in each channel (here the guiding
center $m$) is conserved. Indeed, due to the conservation of
the angular momentum, the forward scattering approximation will be
always restricted to first order effects in $\gamma $, as we
discuss in Sec. (III.C). But, as we stressed above, this is not a
very restrictive condition, because the previous methods of
calculation are not even able to include all these first order
effects.

We saw above that if we formally divide a two dimensional IQH\
system into $m $ interacting one dimensional channels, the strong
magnetic field turns the neglect of the net transfer of particles
between the ``$m$'' channels into a reasonable approximation. So,
this system looks very convenient for the use of a bosonization
technique, once we can skirt the treatment of the transfer of
particles between many different one dimensional channels in this
first approximation. The latter is known to be the main difficulty
with the higher dimensional bosonization methods
\cite{houghton,Castro}. However, despite the problem with the many
channels particle transfer, there are many techniques developed to
study the effects of the net transfer of particles between two
channels in a bosonization method. This problem is analogous to
the inclusion of the backward scattering in the usual bosonization
of a one-dimensional interacting electronic system, which makes
the transfer of particles between the right- and left-moving channels \cite{RefI}.
So, in a future work, we intend to
included the transfer between two neighbors ``$m$'' channels. This
neighbor transfer seems to us a promising alternative expansion
parameter to the interaction strength $\gamma $, as we discuss
further in the conclusion. However, in this work, we will restrict
ourselves to the first order terms in the strong field
approximation or, equivalently, to the forward scattering
approximation.

\subsection{The Bosonic Description}

Motivated by the above discussion, we will introduce bosonic operators
associated with electron-hole excitations that conserve $m$, as in the
bosonization scheme previously developed by two of us \cite{Harry}. If we
extend the Hilbert space of all guiding centers to negative energy values,
in the spirit of the Luttinger model \cite{Lutt,VanD}, we get that the
operator
\begin{equation}
b_{n,m}^{\dagger }=\frac{1}{\sqrt{n}}\sum_{p=-\infty }^{\infty
}c_{n+p,m}^{\dagger }c_{p,m},  \label{b+}
\end{equation}
and its Hermitian conjugate obey exact bosonic relations:
\begin{equation}
\left[ b_{n,m},b_{n^{\prime },m^{\prime }}^{\dagger }\right] =\delta
_{m,m^{\prime }}\delta _{n,n^{\prime }},\qquad \left[ b_{n,m},b_{n^{\prime
},m^{\prime }}\right] =0.  \label{comut1}
\end{equation}
This extension is a well controlled approximation as long as the negative energy states
are kept inert. So, the system's Fermi energy must be much higher than the
characteristic Coulomb interaction, $\nu \hbar \omega _{c}\gg
e^{2}/\varepsilon l$, which is essentially the strong field approximation.

The Hamiltonian (\ref{H0}) can be rewritten in terms of the previously
defined bosonic operators as \cite{VanD}
\begin{equation}
H_{0}=\hbar \omega _{c}\sum_{m=0}^{N_{\Phi }-1}\left[ \sum_{n>0}nb_{n,m}^{%
\dagger }b_{n,m}+\frac{1}{2}\hat{N}_{m}\left( \hat{N}_{m}+1\right) \right] ,
\label{H0b}
\end{equation}
where $\hat{N}_{m}=\sum_{p=-\infty }^{\infty }\left( c_{p,m}^{\dagger
}c_{p,m}-\left\langle G_{\mathbf{0}}\right| c_{p,m}^{\dagger }c_{p,m}\left|
G_{\mathbf{0}}\right\rangle \right) $ is the operator which counts the
number of particles relative to the \textit{reference state} $\left| G_{%
\mathbf{0}}\right\rangle $ in the channel $m$. The last term gives the
ground state energy $\left\langle G_{\mathbf{N}}\right| H_{0}\left| G_{%
\mathbf{N}}\right\rangle $ of a system with a distribution of particles per
channel given by the vector $\mathbf{N=}\left( N_{0},...,N_{\phi -1}\right) $
and the first one describes the excitations above this state. Formally,
what allows us to rewrite (\ref{H0}) in this alternative bosonic
representation, without a further approximation, is the convenient linear
dispersion relation in $n$ of the system \cite{VanD}.

If we now include the Coulomb interaction term (\ref{HI}) in the \textit{%
forward scattering approximation} and restrict ourselves to the spectrum of
low energy excitations (close to the highest filled, $\nu -1$, Landau level)
the total Hamiltonian can be expressed in terms of the bosonic operators $%
b_{n,m}^{\left( \dagger \right) }$, eq.(\ref{b+}). Even restricting to processes
that conserve the total number of electrons in each guiding center, as in the one
dimensional case \cite{Mahan}, there is still more than one way to pair the fermionic operators in
(\ref{HI}) to yield the bosonic bilinear terms. If we set $m_{1}=m_{1}^{\prime }$
and $m_{2}=m_{2}^{\prime }$ in (\ref{HI}) we get the direct scattering
processes whereas exchange scattering processes are obtained for $%
m_{1}=m_{2}^{\prime }$ and $m_{2}=m_{1}^{\prime }$. This leads us to
\begin{equation}
H_{I}=\frac{e^{2}}{\epsilon l}\sum_{n>0,m,m^{\prime }}\left[ g_{di}\left(
n;m,m^{\prime }\right) +g_{ex}\left( n;m,m^{\prime }\right) \right]
b_{n,m}^{\dagger }b_{n,m^{\prime }}+H_{N},  \label{Htot}
\end{equation}
with
\begin{eqnarray}
g_{di}\left( n;m,m^{\prime }\right) &=&n\int d\bar{q}\Phi _{m}\left( \bar{q}%
\right) \Phi _{m^{\prime }}\left( \bar{q}\right) \tilde{V}_{di}^{n}\left(
\bar{q}\right) ,  \label{gdi} \\
g_{ex}\left( n;m,m^{\prime }\right) &=&E_{Self}^{n}\delta _{m,m^{\prime }}
\label{gex} \\
&&-n\int d\bar{q}\Phi _{m}\left( \bar{q}\right) \Phi _{m^{\prime }}\left(
\bar{q}\right) \tilde{V}_{ex}^{n}(\bar{q}\mathbf{)}\,,  \notag
\end{eqnarray}
where $\bar{q}=lq$, $\Phi _{m}\left( \bar{q}\right) =\sqrt{\bar{q}}%
G_{m,m}\left( \frac{\bar{q}^{2}}{2}\right) e^{-\bar{q}^{2}/4}$,
\begin{eqnarray}
\tilde{V}_{di}^{n}\left( \bar{q}\right) &=&\frac{e^{-\bar{q}^{2}/2}}{\bar{q}}%
\left| G_{\nu -1,\nu -1+n}\left( \mathbf{\bar{q}}\right) \right| ^{2},
\label{Vdi} \\
E_{Self}^{n} &=&\int d\bar{r}e^{-\frac{\bar{r}^{2}}{2}}L_{\nu -1}^{1}\left(
\frac{\bar{r}^{2}}{2}\right) \left[ L_{\nu -1}^{0}\left( \frac{\bar{r}^{2}}{2%
}\right) \right.  \notag \\
&&\left. -L_{\nu -1+n}^{0}\left( \frac{\bar{r}^{2}}{2}\right) \right] ,
\label{HautoE} \\
\tilde{V}_{ex}^{n}(\bar{q}) &=&\int d\bar{k}e^{-\frac{\bar{k}^{2}}{2}}L_{\nu
-1+n}(\frac{\bar{k}^{2}}{2})L_{\nu -1}(\frac{\bar{k}^{2}}{2})\mathcal{J}%
_{0}\left( \bar{k}\bar{q}\right) ,  \label{Vexn2}
\end{eqnarray}
$\bar{r}=r/l$ and $\mathcal{J}_{0}\left( x\right) $ is the zero order Bessel
function.

In the above expressions, $g_{di}\left( n;m,m^{\prime }\right) $
is the contribution (see Appendix B for details) from the direct
interaction process \cite{Halp,Kallin}, that describes the
recombination and destruction of the electron-hole pair, and
$g_{ex}\left( n;m,m^{\prime }\right) $ is the contribution from
the exchange process (see Appendix B for details). Moreover,
$g_{ex}\left( n;m,m^{\prime }\right) $ describes the bound state
of a hole in the level $\nu -1$ and an electron in the level $\nu
-1+n$ \ moving along different guiding centers \cite{Halp}. This
is the quantum analogue of the classical interaction of two
particles of opposite charge under a magnetic field, which move
parallel to one another with a constant velocity perpendicular to
their separation. This term has a constant diagonal contribution
$E_{Self}^{n}$ that is the difference between the exchange
self-energy of an electron in the in the excited level $\nu -1+n$
and the level $\nu -1$ from which it was removed. Besides, we see
that the forward scattering approximation does not generate a
diagonal interaction in the guiding centers $m$.

The term $H_{N}$ in (\ref{Htot}) contains only terms dependent on
the number operator. Since they commute with the bosonic operators
they have no influence on the system´s dynamics but only on its
chemical potential (see section \ref{Che}).

\subsection{Diagonalization of the total Hamiltonian\label{Dia}}

The total Hamiltonian in the forward scattering approximation,
\begin{eqnarray}
H_{T} &=&H_{0}+H_{I}  \notag \\
&=&\hbar \omega _{c}\sum_{n>0}\sum_{m}nb_{n,m}^{\dagger }b_{n,m}
\label{Htotb} \\
&&+\frac{e^{2}}{\epsilon l}\sum_{n>0}\sum_{m,m^{\prime }}g\left(
n;m,m^{\prime }\right) b_{n,m}^{\dagger }b_{n,m^{\prime }}+H_{N^{\prime }},
\notag
\end{eqnarray}
with $g\left( n;m,m^{\prime }\right) =g_{di}\left( n;m,m^{\prime }\right)
+g_{ex}\left( n;m,m^{\prime }\right) $, can be diagonalized by an
appropriate linear combination of guiding center bosonic operators $%
b_{n,m}^{\dagger }$. It is easy to see that such a combination is given by
\begin{equation}
b_{n}^{\dagger }\left( \bar{q}\right) =\sum_{m=0}^{N_{\Phi }-1}\Phi
_{m}\left( \bar{q}\right) b_{n,m}^{\dagger },\quad \Phi _{m}\left( \bar{q}%
\right) =\sqrt{\bar{q}}L_{m}\left( \frac{\bar{q}^{2}}{2}\right) e^{-\bar{q}%
^{2}/4}.  \label{b1}
\end{equation}
This new representation describes the degeneracy breaking between
excitations like (a) and (d) in Fig.(\ref{Fig1}). By using the completeness
and orthogonality relations of the $\Phi _{m}(q)$ functions,
\begin{eqnarray}
\sum_{m=0}^{N_{\phi }-1}\Phi _{m}\left( \bar{q}\right) \Phi _{m}\left( \bar{q%
}^{\prime }\right) &=&\delta \left( \bar{q}-\bar{q}^{\prime }\right) ,
\notag \\
\int_{0}^{\infty }d\bar{q}\Phi _{m}\left( \bar{q}\right) \Phi _{m^{\prime
}}\left( \bar{q}\right) &=&\delta _{m,m^{\prime }},  \label{Prof1}
\end{eqnarray}
we see that the operators $b_{n}\left( \bar{q}\right) $ also obey bosonic
commutations relations
\begin{equation}
\left[ b_{n}\left( \bar{q}\right) ,b_{n^{\prime }}^{\dagger }\left( \bar{q}%
^{\prime }\right) \right] =\delta _{n,n^{\prime }}\delta \left( \bar{q}-\bar{%
q}^{\prime }\right) ,\quad \left[ b_{n}\left( \bar{q}\right) ,b_{n^{\prime
}}\left( \bar{q}^{\prime }\right) \right] =0,  \label{b5}
\end{equation}
and are related to the $b_{n,m}^{\dagger }$ bosons by
\begin{equation}
b_{n,m}^{\dagger }=\int_{0}^{\infty }d\bar{q}\Phi _{m}\left( \bar{q}\right)
b_{n}^{\dagger }\left( \bar{q}\right) .  \label{bnmq}
\end{equation}
The same holds for the operator $\hat{N}\left( \bar{q}\right) $, which plays
the role of a {}``\textit{zero-mode boson}'':
\begin{equation}
\hat{N}\left( \bar{q}\right) =\sum_{m=0}^{N_{\phi }-1}\Phi _{m}\left( \bar{q}%
\right) \hat{N}_{m}.  \label{b7}
\end{equation}

The total Hamiltonian can then be directly rewritten in this new
representation as
\begin{equation}
H_{T}=\sum_{n>0}\int d\bar{q}E_{n}\left( \bar{q}\right) b_{n}^{\dagger
}\left( \bar{q}\right) b_{n}\left( \bar{q}\right) +H_{N^{\prime }},
\label{HTh}
\end{equation}
with the energy spectrum
\begin{equation}
E_{n}\left( \bar{q}\right) =n\left[ \hbar \omega _{c}+\tilde{V}_{ex}^{n}(%
\bar{q}\mathbf{)+}\tilde{V}_{di}^{n}\left( \bar{q}\right) \right]
+E_{Self}^{n}.  \label{Enq}
\end{equation}

The fact that the Hamiltonian is diagonal in the basis $b_{n}\left( \bar{q}%
\right) $ is a consequence of the forward scattering approximation. In the
symmetric gauge, the system's angular momentum is given by $L=\sum_{i}\left(
n_{i}-m_{i}\right) $ and is a constant of motion. Therefore, in (\ref{HI}),
the processes with non-zero matrix elements are the ones at which
\begin{equation}
\left( n_{1}^{\prime }-m_{1}^{\prime }\right) +\left( n_{2}^{\prime
}-m_{2}^{\prime }\right) =\left( n_{1}-m_{1}\right) +\left(
n_{2}-m_{2}\right) .  \label{CM2}
\end{equation}
In the forward scattering approximation we have $m_{1}=m_{1}^{\prime }$ and $%
m_{2}=m_{2}^{\prime }$ (or $m_{1}=m_{2}^{\prime }$ and $m_{2}=m_{1}^{\prime
} $, in the exchange processes). Eq.(\ref{CM2}) is then reduced to $%
\Delta n_{1}+\Delta n_{2}=0$, with $\Delta n_{i}=n_{i}^{\prime
}-n_{i}$. So, we are restricted to the first order terms in a
degenerate perturbation theory, which present only terms that
conserve the energy associated with the free portion, $H_{0}$, of
the total Hamiltonian. In the bosonic representation this means
terms like $b_{n}^{\dagger }\left( \bar{q}\right)
b_{n}\left( \bar{q}\right) $, but neither like $b_{n}^{\dagger }\left( \bar{q%
}\right) b_{n^{\prime }}\left( \bar{q}\right) $, with $n\neq n^{\prime }$,
nor like $b_{n}^{\left( \dagger \right) }\left( \bar{q}\right) b_{n^{\prime
}}^{\left( \dagger \right) }\left( \bar{q}\right) $. Formally this form
appears in our expressions after the angular integration over $\arg \left(
\mathbf{q}\right) $ in the matrix elements (see\ Appendix B), which clearly
imposes the angular momentum conservation.

\section{Comparison with the Tomonaga-Luttinger Model and the 2D-Bosonization%
\label{Ground}}

\begin{figure}[ptb]
 \includegraphics[width=1.0%
\columnwidth,keepaspectratio]{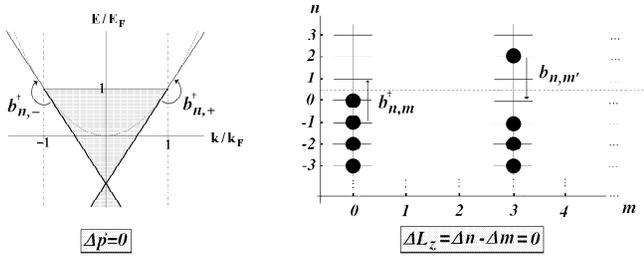}
\par
 \caption{(a) Illustration of the linear momentum conservation by the action
of $b_{n,+}^{\dagger}b_{n,-}^{\dagger}$ in the Tomonaga-Luttinger model. (b)
Illustration of the angular momentum conservation by the action of the $%
b_{n,m}^{\dagger}b_{n,m^{\prime}}$ in our model. The particles are kept in
their channels, therefore $\Delta m=0$, and the variation on $n$ $,\Delta n$%
, must be equal to zero too.}
\label{Fig2}
\end{figure}

We can compare our Hamiltonian (\ref{Htotb}) with the one of the
Tomonaga-Luttinger model \cite{voit,Schon}:
\begin{eqnarray}
H &=&\sum_{n>0}k_{n}\left\{ \sum_{j=\pm }\left[ \hbar v_{F}+g_{4}\left(
k_{n}\right) \right] b_{n,j}^{\dagger }b_{n,j}\right.  \notag \\
&&\left. +\sum_{j\neq j^{\prime }}g_{2}\left( k_{n}\right) \left(
b_{n,j}^{\dagger }b_{n,j^{\prime }}^{\dagger }+b_{n,j}b_{n,j^{\prime
}}\right) \right\} +H_{N},  \label{Htomo}
\end{eqnarray}
associated with one-dimensional (1D) fermion systems. In this model the
system's neutral excitations are distributed in two momentum channels, $%
j=\pm $ (left and right-moving fermions), presenting energies given by the
linearization of the dispersion relation close to the two Fermi points, $\pm
k_{F}$ (see Fig.(\ref{Fig2})). In our two dimensional model the neutral
excitations are distributed among the $m=N_{\phi }$ guiding centers.
Comparing (\ref{Htotb}) and (\ref{Htomo}) we see that, despite the different
number of channels, we have the equivalence relations:
\begin{equation*}
g_{4}\left( k_{n}\right) \rightarrow g\left( n;m,m\right) ,\qquad \qquad
g_{2}\left( k_{n}\right) \rightarrow g\left( n;m,m^{\prime }\right) .
\end{equation*}

In the Tomonaga-Luttinger model the terms associated with the interaction of
particles in different channels, described by $g_{2}\left( k_{n}\right) $,
present the form $b_{n,j}^{\dagger }b_{n,j^{\prime }}^{\dagger }$ or $%
b_{n,j}b_{n,j^{\prime }}$. This is because of the linear momentum
conservation
in the system: $b_{n,+}^{\dagger }$ creates an excitation with wave vector $%
k_{F}+k_{n}$, while $b_{n,-}^{\dagger }$ creates an excitation with wave
vector $-k_{F}-k_{n}$. So, $b_{n,j}^{\dagger }b_{n,j^{\prime }}^{\dagger }$
creates an excitation with total momentum equal to zero (see Fig.(\ref{Fig2}%
)).

In the Tomonaga-Luttinger model this portion of the Hamiltonian, associated
with $g_{2}\left( k_{n}\right) $, does not commute with the kinetic energy
term $H_{0}=\hbar v_{F}\sum_{n>0}\sum_{j=\pm }k_{n}b_{n,j}^{\dagger }b_{n,j}$%
. Therefore, it modifies the system´s ground state, which has a
distribution of electron-hole pairs. The terms associated with
$g_{4}\left( k_{n}\right)$ describe the interaction between
particles in the same channel. These terms commute with $H_{0}$
and their influence is limited to removing degeneracies in the
excitations \cite{voit}. In our model, all the terms associated
with the interparticle interaction, in the same or different
channels, present the same form $b_{n,m}^{\dagger }b_{n,m^{\prime
}}$. As we have seen in the section above (after having made the
forward scattering approximation) this form results from the
angular momentum conservation, which does not allow the presence
of
terms like $b_{n,m}^{\dagger }b_{n,m^{\prime }}^{\dagger }$ or $%
b_{n,m}b_{n,m^{\prime }}$ (see Fig.(\ref{Fig2})). Therefore, we will have a
ground state like a Fermi sea, without particle-hole excitations or Landau
levels mixing. All the interactions $g\left( n;m,m^{\prime }\right) $, with $%
m=m^{\prime }$ or $m\neq m^{\prime }$, contribute to removing the
degeneracies of the $n$ levels.

In a bosonized Hamiltonian, the presence of terms like $b^{\dagger
}b^{\dagger }$ and $bb$, which separately do not conserve energy,
``spreads'' the single-particle spectral functions and gives rise
to the \textit{anomalous dimension}, $\alpha $, in the
correlations functions \cite {Voit2}. In the spin-full
Tomonaga-Luttinger model, besides the spin-charge separation
effect, the presence of this anomalous dimension results in the
absence of fermionic quasi-particles (vanishing of
the quasi-particle residue $z_{k}\sim \left| k-k_{F}\right| ^{\alpha }$ as $%
k\rightarrow k_{F}$). In other words in the non-Fermi liquid
behavior of the model. In the 2D bosonization of the electron gas
\cite{Castro} terms like $b^{\dagger }b^{\dagger }$ and $bb$,
although present in the original interaction Hamiltonian, can be
shown to vanish in the thermodynamic limit and in the asymptotic
low-energy regime, giving rise to the Fermi liquid behavior. In
our model it is clear that we will not have effects coming from an
anomalous dimension, but we will have other effects that will give
rise to a non-normal behavior.

\section{Including the Electron Spin}

\subsection{The Free Hamiltonian}

Including the electron spin, the free Hamiltonian (\ref{H0}) becomes:
\begin{equation}
H_{0}=\hbar\omega_{c}\sum_{s=\pm}\sum_{m=0}^{N_{\Phi}-1}\sum_{n=0}^{%
\infty}(n+1/2)c_{n,m,s}^{\dagger}c_{n,m,s}.  \label{H0S}
\end{equation}
Considering the extension of the Hilbert space to negative energy values, we
can write $H_{0}$ in terms of the new bosonic operators, $b_{n,s}\left(\bar{q%
}\right)$:
\begin{equation}
H_{0}=\hbar\omega_{c}\sum_{s=\pm}\int d\bar{q}\left[\sum_{n>0}nb_{n,s}^{%
\dagger}\left(\bar{q}\right)b_{n,s}\left(\bar{q}\right)+\frac{1}{2}\hat{N}%
_{s}^{2}\left(\bar{q}\right)\right],  \label{H0S2}
\end{equation}
\begin{equation}
b_{n,s}\left(\bar{q}\right)=\sum_{m=1}^{N_{\phi}-1}\Phi_{m}\left(\bar{q}%
\right)\frac{1}{\sqrt{n}}\sum_{p=-\infty}^{\infty}c_{p,m,s}^{%
\dagger}c_{p+n,m,s}.  \label{bnms}
\end{equation}

However, it is more convenient to define the following usual bosonic
operators \cite{voit}:
\begin{eqnarray}
b_{n}\left( \bar{q}\right) _{\rho } &=&\frac{1}{\sqrt{2}}\left[
b_{n,+}\left( \bar{q}\right) +b_{n,-}\left( \bar{q}\right) \right] ,
\label{OBCS1} \\
b_{n}\left( \bar{q}\right) _{\sigma } &=&\frac{1}{\sqrt{2}}\left[
b_{n,+}\left( \bar{q}\right) -b_{n,-}\left( \bar{q}\right) \right] ,
\label{OBCS2}
\end{eqnarray}
which are, as we will see, directly related to the charge and spin
fluctuations in our system. Writing (\ref{H0S2}) in terms of these
operators, we have:
\begin{equation}
H_{0}=\hbar \omega _{c}\sum_{i=\rho ,\sigma }\int d\bar{q}\left[
\sum_{n>0}nb_{n}^{\dagger }\left( \bar{q}\right) _{i}b_{n}\left( \bar{q}%
\right) _{i}+\frac{1}{2}\hat{N}^{2}\left( \bar{q}\right) _{i}\right] .
\label{H0SC2}
\end{equation}

The bosonic operators $b_{n}\left( q\right) _{\rho }$ and $b_{n}\left(
q\right) _{\sigma }$ commute. Therefore, they are associated with two
different set of excitations. In one dimensional systems they describe the
independent charge and spin fluctuations \cite{voit,LuEm,Meden}. In our
system these operators play an equivalent role. If the higher Landau levels
have both spin sublevels filled (see the next section) the system ground
state will present total spin $\mathbf{S=0}$. Therefore, we will have
triplet or singlet excitations. The triplet excitations, described by the
operators $b_{n}^{\left( \dagger \right) }\left( \bar{q}\right) _{\rho }$,
give rise to poles in the density response function $\chi _{\rho }\left(
q,\omega \right) $ whereas the singlet ones, described by the operators $%
b_{n}^{\left( \dagger \right) }\left( \bar{q}\right) _{\sigma },$ to poles
in the spin-density response function $\chi _{\sigma }\left( q,\omega
\right) $ \cite{Halp}.

\subsection{The Zeeman Energy\label{Zem}}

The inclusion of the electron spin brings together the Zeeman energy, $%
H_{Ze}=\frac{\hbar }{2}\mu _{B}gB\sum_{s=\pm
}\sum_{n,m}sc_{n,m,s}^{\dagger }c_{n,m,s}$, which splits each
Landau level into two sublevels  with energy separation of the
order of $\Delta E_{z}=\frac{\hbar }{2}\mu _{B}gB$. However, we
will
treat only excitations between Landau levels. Our bosonic operators $%
b_{n}\left( \bar{q}\right) _{\rho }$ or $b_{n}\left( \bar{q}\right) _{\sigma
}$ are associated with particle-hole pairs with the same spin, see eqs. (\ref
{bnms}-\ref{OBCS2}). These are the only possible excitations for a system
with both spin sublevels filled $\left( \mathbf{S}=0\right) $. The particle'%
s spin will be conserved in the transitions and therefore the Zeeman energy
will not have any influence in the system dynamics. One can easily
see that $H_{Ze}$ commutes with $b_{n}\left( \bar{q}\right) _{\rho }$ and $%
b_{n}\left( \bar{q}\right) _{\sigma }$. An appropriate description
of the intra-Landau levels excitations \cite{Longo}, through an
appropriate bosonization method, was also presented recently by
one of us \cite{Dor}.

\subsection{The Interacting Hamiltonian}

We saw that the inclusion of the spin and the definition of the operators (%
\ref{OBCS1}-\ref{OBCS2}) bring the free Hamiltonian $H_{0}$ (\ref{H0S}) to
the form (\ref{H0SC2}). Now, once the exchange processes occur only between
particles with the same spin, the exchange Hamiltonian $H_{I}^{Ex}$, (\ref
{HIEX}), will be diagonal in the $s$ index, like $H_{0}$, (\ref{H0S}).
Therefore, it will be transformed in the same way as $H_{0}$, being written
in terms of $b_{n}\left( \bar{q}\right) _{\rho }$ and $b_{n}\left( \bar{q}%
\right) _{\sigma }$. On the other hand, the direct interaction $H_{I}^{Dir}$%
, (\ref{hdirro2}), will not be diagonal in the spin index. The sum
over the spin indices, $\sum_{s,s^{\prime }}$, results in the
cancellation of the terms proportional to $b_{n}\left(
\bar{q}\right) _{\sigma }$ and in the presence of an interaction
equal to $2\tilde{V}_{di}^{n}\left( \bar{q}\right) $ associated
with the charge operators $b_{n}\left( \bar{q}\right) _{\rho }.$
Physically, the direct interaction gives rise only to charge
excitations, while the exchange interaction gives rise to charge
and spin excitations. The final Hamiltonian will then present the
charge and spin fluctuations separated:
\begin{eqnarray}
H_{T} &=&H^{\left( \rho \right) }+H^{\left( \sigma \right) },  \label{HTTS}
\\
H^{\left( i\right) } &=&\sum_{n>0}\int d\bar{q}E_{n}^{\left( i\right)
}\left( \bar{q}\right) b_{n}^{\dagger }\left( \bar{q}\right) _{i}b_{n}\left(
\bar{q}\right) _{i},  \notag \\
E_{n}^{\left( i\right) }\left( \bar{q}\right) &=&n\left[ \hbar \omega
_{c}+2\delta _{i,\rho }\tilde{V}_{di}^{n}\left( \bar{q}\right) +\tilde{V}%
_{ex}^{n}(\bar{q}\mathbf{)}\right] +E_{Self}^{n},  \label{HTS}
\end{eqnarray}
with $i=\rho ,\sigma $ and the terms of the energy spectrum $E_{n}^{\left(
i\right) }\left( \bar{q}\right) $ given in (\ref{Vdi}-\ref{Vexn2}). We did
not write the terms proportional to the number operators. In the limit $%
q\rightarrow \infty $ we will have the asymptotic value $E_{n}^{\left(
i\right) }=n\hbar \omega _{c}+E_{Self}^{n}$ since $\tilde{V}_{ex}^{n}(%
\bar{q}\mathbf{)}$ and $\tilde{V}_{di}^{n}\left( \bar{q}\right) $ goes to zero.

\section{Spectrum of Neutral Excitations}

For the first excited level, $n=1$, the neutral excitation spectrum that we have obtained
, eq. (\ref{HTS}), is identical to the one determined in eq.(4.2) of the
Kallin and Halperin's paper \cite{Halp}. We just have to perform the angular
integration over $\arg \left( \mathbf{r}\right) $ in the latter to obtain (%
\ref{HTS}). For $n>1$ the only difference is the $n$ factor multiplying $%
\tilde{V}_{di}^{n}\left( \bar{q}\right) $ and $\tilde{V}_{ex}^{n}(\bar{q})$,
that we will discuss below. The authors of Ref.[4] clarify that (\ref
{HTS}), within the strong field approximation, applies for two cases, at
which it is exact to first order in $\left( e^{2}/\epsilon l\right) /\hbar
\omega _{c}$: (i) excitations associated with $n=1$ in systems with
arbitrary $\nu $ and (ii) excitations associated with arbitrary $n$ since
the system presents $\nu =1$; in both cases the system must present the two
spin Landau sublevels filled, as we suppose in sec.(\ref{Zem}). Despite
these restrictions, these are the excitations of major interest. Outside the
domain of these conditions eq.(\ref{HTS}) represents a less controlled
approximation, once the multi-exciton states were neglected.

In the bosonization procedure we can easily understand the origin of these
restrictions. To obtain a precise bosonic description of the excitations we
have approximated all the matrix elements $G_{p,p+n}$, that describe
the excitation of an electron to the $p+n$ level, leaving a hole in the $p$
level, by the Fermi level matrix element, $\left. G_{p,p+n}\right| _{p=\nu
-1}$= $G_{\nu -1,\nu -1+n}$, see Appendix B. In this case we have a precise
description for the excitations associated with $n=1$, as they are
represented by the unique element $G_{\nu -1,\nu }$, and for the arbitrary
excitations in a system presenting $\nu =1$, since they are all
associated with the same element $G_{0,n}$. In these two cases all the
particle-hole pairs that contribute to a given excitation are associated
with the same Landau levels. So, we do not have to consider all the $n$
Landau levels between $\nu -n$ and $\nu -1$ that would contribute to a given
$E_{n}\left( q\right) $ if $\nu -n>0$ \cite{Halp}$.$ However, once we
introduce negative energy states, even for filling $\nu =1$ we will have a
contribution of $n$ states. This is the origin of the $n$ factor multiplying
$\tilde{V}_{di}^{n}\left( \bar{q}\right) $ and $\tilde{V}_{ex}^{n}(\bar{q})$
in (\ref{HTS}). But, as we will see, the $n=1$ magnetoplasma mode will be
the most import excitation in our analysis and, in the strong field limit $%
\nu \hbar \omega _{c}\gg e^{2}/\epsilon l,$ the higher magnetoplasmon modes
will not present a relevant modification due this $n$ factor.

We justify the approximation in the matrix elements, $\left.
G_{p,p+n}\right| _{p=\nu -1}=G_{\nu -1,\nu -1+n}$, with the strong field
approximation, see Appendix B. However, we should mention that it is also a
good approximation to describe the low energy excitations $\left( n\ll \nu
\right) $ of a system in low magnetic field $\left( \nu \gg 1\right) $. In
this case, for small wave vectors $\bar{q}$, it can be shown that the matrix
elements $G_{\nu -1,\nu -1+n}\left( \bar{q}\right) $ depend approximately
only on $n$ and the value adopted by $\nu $ is irrelevant. This property was
explored in a similar bosonization method \cite{Harry}.

The dispersion curves for the modes close to $\hbar \omega _{c}$ are
obtained calculating (\ref{HTS}) for $n=1.$\ For $\nu =1$ \cite{Obs}, e.g.,
we got:
\begin{eqnarray}
E_{\nu =n=1}^{\left( i\right) }\left( \bar{q}\right) &=&\hbar \omega _{c}+%
\frac{e^{2}}{\epsilon l}\frac{1}{2}\sqrt{\frac{\pi }{2}}\left\{ 1-e^{-\frac{%
\bar{q}^{2}}{4}}\left[ \left( 1+\frac{\bar{q}^{2}}{2}\right) I_{0}\left(
\frac{\bar{q}^{2}}{4}\right) \right. \right.  \notag \\
&&\left. \left. -\frac{\bar{q}^{2}}{2}I_{1}\left( \frac{\bar{q}^{2}}{4}%
\right) \right] +2\delta _{i,\rho }\sqrt{\frac{2}{\pi }}\bar{q}e^{-\frac{%
\bar{q}^{2}}{2}}\right\} .  \label{En1}
\end{eqnarray}
These expressions, for the charge and spin modes, $i=\rho ,\sigma $, are
exactly the eqs. (4.4) and (4.8) in Ref.[4] or (3.3) in Ref.[31]. The
first term between curly brackets is independent of $q$ and corresponds to $%
E_{Self}^{n=1}$. The second term corresponds to
$\tilde{V}_{ex}^{n=1}(\bar{q}\mathbf{)}$
and the last one to $2\delta _{i,\rho }\tilde{V}_{di}^{n=1}\left( \bar{q}%
\right) $. For higher $n$ or $\nu $ values we have much longer expressions,
so we do not explicitly show them here.

\section{Single-Particle Properties}

\subsection{Fermionic Operator\label{Che}}

The bosonic representation of fermionic fields, also known as the \textit{%
Mattis-Mandelstam} formula, is an identity in one dimensional systems if the
theory can be formulated in terms of a set of fermionic operators, $\left\{
c_{k,j},c_{k^{\prime },j^{\prime }}^{\dagger }\right\} =\delta _{j,j^{\prime
}}\delta _{k,k^{\prime }}$, which are labeled by the ``species'' index $%
j=1,....,N_{j}$ and a discrete and unbounded ``momentum'' index $k,$ which
typically labels the eingenenergies, $E_{k}$, of the non-interacting system
\cite{VanD}. The fermionic fields, $\psi _{j}\left( x\right) $, are defined
as:
\begin{equation}
\psi _{j}\left( x\right) \equiv \sqrt{\frac{2\pi }{L}}\sum_{n=-\infty
}^{\infty }e^{-ik_{n}x}c_{n,j},  \label{OF}
\end{equation}
where $L$ is the length associated with the system size and $c_{n,j}$ remove
the fermion from the state $k_{n}=\frac{2\pi }{L}n$. They have the
following bosonic representation \cite{VanD,haldane2}:
\begin{eqnarray}
\psi _{j}\left( x\right) &=&\frac{1}{\sqrt{\epsilon }}U_{j}e^{-i\frac{2\pi }{%
L}\hat{N}_{j}x}e^{-i\Phi _{j}\left( x\right) },  \label{OFL} \\
\Phi _{j}\left( x\right) &=&i\sum_{k_{n}>0}\frac{e^{-\epsilon k_{n}/2}}{%
\sqrt{n}}\left( e^{-ik_{n}x}b_{n,j}-e^{ik_{n}x}b_{n,j}^{\dagger }\right) ,
\label{OFB3} \\
b_{n,j} &=&\frac{1}{\sqrt{n}}\sum_{p=-\infty }^{\infty }c_{p-n,j}^{\dagger
}c_{p,j},  \label{bj}
\end{eqnarray}
where $b_{n,j}$ are bosonic operators \cite{VanD} and $\epsilon >0$ is an
infinitesimal regularization parameter such that $\epsilon \rightarrow 0^{+}$
at the end of the calculations.

We get a fermionic field appropriate for a bosonic representation with $%
b_{n,m}$, introducing the phase representation \cite{Harry}
\begin{equation}
\psi _{m}\left( \theta \right) \equiv \frac{1}{\sqrt{2\pi }}\sum_{n=-\infty
}^{\infty }e^{-in\theta }c_{n,m}.  \label{OFT}
\end{equation}
In this way the set of operators $\left\{ c_{n,m},b_{n,m},\hat{N}_{m}\text{,}%
\psi _{m}\left( \theta \right) \right\} $ is completely analogous to the set
$\left\{ c_{k_{n},j},b_{n,j}\text{,}\hat{N}_{j},\psi _{j}\left( x\right)
\right\} $ if we make the replacements: $j\rightarrow m$, $x\rightarrow
\theta $, $k_{n}\rightarrow n$, $\frac{2\pi }{L}\hat{N}_{j}\rightarrow \hat{N%
}_{m}$ and $\sqrt{\frac{2\pi }{L}}\rightarrow \frac{1}{\sqrt{2\pi }}$.
Therefore, writing the fermionic field $\psi _{m}\left( \theta \right) $ in
terms of the operators $b_{n}\left( \bar{q}\right) $ of (\ref{b1}), it
presents the following representation, analogous to (\ref{OFL}-\ref{bj}):
\begin{eqnarray}
\psi _{m}\left( \theta \right) &=&\lim_{\epsilon \rightarrow 0}\frac{1}{%
\sqrt{2\pi \epsilon }}U_{m}e^{-i\hat{N}_{m}\theta }e^{-i\Theta _{m}\left(
\theta \right) },  \label{OFE} \\
\Theta _{m}\left( \theta \right) &=&i\int_{0}^{\infty }d\bar{q}\Phi
_{m}\left( \bar{q}\right) \sum_{n=1}^{\infty }\frac{e^{-\frac{n\epsilon }{2}}%
}{\sqrt{n}}\left[ e^{-in\theta }b_{n}\left( \bar{q}\right) \right.  \notag \\
&&\left. -e^{in\theta }b_{n}^{\dagger }\left( \bar{q}\right) \right]
\label{FaseT}
\end{eqnarray}
In order to determine the time evolution of the fermion field (\ref{OFE}) we
first note that $b_{n}\left( \bar{q},t\right) =e^{-iE_{n}\left( \bar{q}%
\right) t}b_{n}\left( \bar{q}\right) $, with $E_{n}\left( \bar{q}\right) $
given by (\ref{Enq}), and that the time evolution of $U_{m}$ is given by
\begin{equation}
U_{m}\left( t\right) =e^{iH_{N}^{m}t}U_{m}e^{-iH_{N}^{m}t}=e^{i\left(
H_{N}^{m}-\left. H_{N}^{m}\right| _{N_{m}+1}\right) t}U_{m}  \label{PotQ}
\end{equation}
where $H_{N}^{m}$ is the portion of the total Hamiltonian proportional to
the $\hat{N}_{m}$ operator. Acting on the ground state we get $\left\langle
0\right| U_{m}\left( t\right) =e^{-i\mu _{m}^{+}t}\left\langle 0\right|
U_{m} $, where $\mu _{m}^{+}$ corresponds to the chemical potential
associated with the creation of a particle in the $m$th channel, $\mu
_{m}^{+}=E_{0}\left( N_{m}+1\right) -E_{0}\left( N_{m}\right) $. Grouping
the terms proportional to $N$ in (\ref{Htotb}), which are given in the
Appendix B, we get
\begin{equation}
\mu _{m}^{+}=\hbar \omega _{c}-\frac{1}{2}V\left( r\mathbf{=}0\right) +\frac{%
1}{2N_{\Phi }}\frac{\tilde{V}\left( q\mathbf{=}0\right) }{2\pi l^{2}}=\mu
^{+}.  \label{PotQ2}
\end{equation}

\subsection{Retarded Green's Functions at $T=0$}

The retarded Green's function, at temperature $T=0$, associated with a
particle in the quantum state $\left| n,m\right\rangle $ (with $n$ measured
from the Fermi level, see Fig.(\ref{Fig1})) is given by \cite{Mahan}:
\begin{eqnarray}
iG_{n,m}^{Ret}\left( t\right) &=&\theta \left( t\right) \left\langle
c_{n,m}\left( t\right) c_{n,m}^{\dagger }\left( 0\right) +c_{n,m}^{\dagger
}\left( 0\right) c_{n,m}\left( t\right) \right\rangle _{0}  \notag \\
&=&i\theta \left( t\right) \int_{0}^{2\pi }d\theta e^{in\theta }\left[
G_{m,\theta }^{>}\left( t\right) +G_{m,\theta }^{<}\left( t\right) \right] ,
\label{GnmT2} \\
iG_{m,\theta }^{>}\left( t\right) &=&\left\langle \psi _{m}\left( \theta
,t\right) \psi _{m}^{\dagger }\left( 0,0\right) \right\rangle _{0},  \notag
\\
iG_{m,\theta }^{<}\left( t\right) &=&\left\langle \psi _{m}^{\dagger }\left(
0,0\right) \psi _{m}\left( \theta ,t\right) \right\rangle _{0},  \notag
\end{eqnarray}
where we use the relation (\ref{OFT}) between the fields $\psi _{m}\left(
\theta \right) $ and $c_{n,m}$. The calculation of $G_{m,\theta }^{\gtrless
}\left( t\right) $, from the representation (\ref{OFE}), is presented in
Appendix C. We get:
\begin{eqnarray}
iG_{n,m}^{Ret}\left( t\right) &=&\theta \left( t\right) \int_{0}^{2\pi
}d\theta e^{in\theta }\left[ e^{-i\mu ^{+}t}e^{-i\theta }K_{m,\theta }\left(
t\right) \right.  \notag \\
&&\left. +e^{-i\mu ^{-}t}K_{m,\theta }^{\ast }\left( t\right) \right]
\label{GretT} \\
K_{m,\theta }\left( t\right) &=&\frac{1}{2\pi }e^{\sum_{n=1}^{\infty }\frac{%
e^{-n\epsilon }}{n}e^{-in\theta }\int_{0}^{\infty }d\bar{q}\Phi
_{m}^{2}\left( \bar{q}\right) e^{-iE_{n}\left( \bar{q}\right) t}},
\label{Kmt}
\end{eqnarray}
where $\mu ^{-}$ corresponds to the chemical potential associated with the
removal of a system's particle.

We can calculate the retarded Green's function associated with different
states. For a particle in a position $\mathbf{r}$ we have
\begin{eqnarray}
iG^{Ret}\left( \mathbf{r},t\right) &=&\theta \left( t\right) \left\langle
\psi \left( \mathbf{r},t\right) \psi ^{\dagger }\left( \mathbf{r},0\right)
+\psi ^{\dagger }\left( \mathbf{r},0\right) \psi \left( \mathbf{r},t\right)
\right\rangle  \notag \\
&=&i\sum_{n^{\prime }=0}^{\infty }\sum_{m=0}^{N_{\phi }-1}\left| \langle
\mathbf{r}|n^{\prime },m\rangle \right| ^{2}\tilde{G}_{n^{\prime
},m}^{Ret}\left( t\right) .  \label{Gretr2}
\end{eqnarray}
where we use the representation of $\psi \left( \mathbf{r}\right) $ in the
Landau level basis, (\ref{expferm}), and $\langle \mathbf{r}|n^{\prime
},m\rangle $ is given by (\ref{G}).\ Notice that we defined $%
G_{n,m}^{Ret}\left( t\right) $, eq. (\ref{GretT}), with the index $n$
relative to the Fermi level, according to Figs.(\ref{Fig1}) and (\ref{FigAw1}%
), whereas the index $n^{\prime }$ above is relative to the single
particle
ground state. So we must write $n^{\prime }=n-\left( \nu -1\right) $ in $%
\tilde{G}_{n^{\prime },m}^{Ret}\left( t\right) $, getting
\begin{equation}
G^{Ret}\left( \mathbf{r},t\right) =\sum_{n=0}^{\infty }\sum_{m=0}^{N_{\phi
}-1}\left| \langle \mathbf{r}|n,m\rangle \right| ^{2}G_{n-\left( \nu
-1\right) ,m}^{Ret}\left( t\right) .  \label{Gretr}
\end{equation}

If we want the Green's function associated with a particle in a different
state, e.g., the wave vector $\mathbf{q}$, we have to replace $\langle
\mathbf{r}|n,m\rangle $ by $\langle \mathbf{q}|n,m\rangle .$

\subsection{Single-Particle Spectral Functions}

\begin{figure}[ptb]
 \includegraphics[width=0.75%
\columnwidth,keepaspectratio]{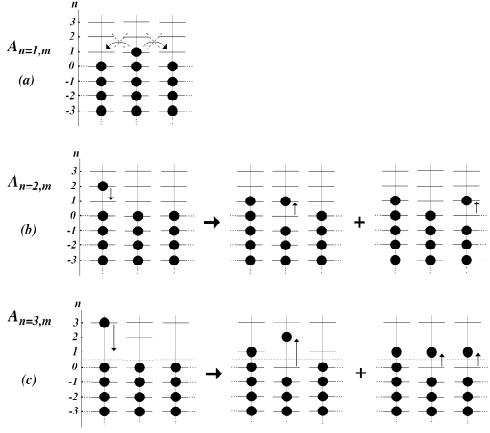}
\par
 \caption{Scheme representing the distribution, in the bosonic modes, of the
energy of the particle introduced in the system:\ (a) a particle created in
the $n=1$ level can not excite any bosonic mode, (b) created in the level $%
n=2$ can excite the bosonic modes associated with $n=1$, $E_{n=1}\left( \bar{%
q}\right) $, (c) created in the $n=3$ can excite the bosonic modes
associated with $n=1$, $E_{n=1}\left( \bar{q}\right) $, and $n=2$, $%
E_{n=2}\left( \bar{q}\right) $, and so on.}
\label{FigAw1}
\end{figure}

\begin{figure}[tbp]
 \includegraphics[width=1%
\columnwidth,keepaspectratio]{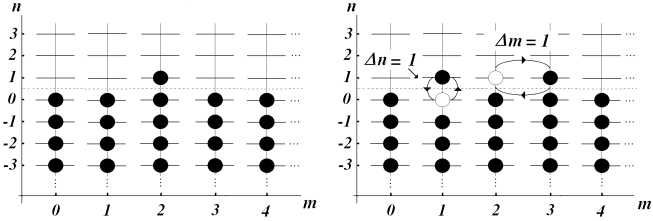}
\par
 \caption{Scheme representing the ``difficulty'' (impossibility in first
order) of a particle added to the system in the first excited level (a) to
make a transition to the neighbor channel, with $\Delta m=+1$ (b). To
conserve angular momentum, $\Delta L_{z}=\Delta n-\Delta m=0$, we must have
another particle making a transition $\Delta n=+1$ simultaneously. This must
be a virtual transition, once it does not conserve the energy.}
\label{Jump}
\end{figure}

Now we will determine the single-particle spectral functions, $A\left(
\omega \right) $, which gives the spectral distribution generated by the
inclusion or removal of a particle in the system. The spectral function
associated with a particle in the sate $\left| n,m\right\rangle $ is given
by
\begin{equation}
A_{n,m}\left( \omega \right) =-\frac{1}{\pi }ImG_{n,m}^{Ret}\left( \omega
\right) =-\frac{1}{\pi }Im\int_{-\infty }^{\infty }dte^{-i\omega
t}G_{n,m}^{Ret}\left( t\right) .  \label{Anm}
\end{equation}
We need the Fourier transform of the function $G_{n,m}^{Ret}\left( t\right) $%
. We then take $n>0$, creating a particle or a hole above the Fermi level
(see Fig.(\ref{Fig1})). Since our ground state does not present electrons
above the Fermi level, see section (\ref{Ground}), only the first term
between square brackets of (\ref{GretT}) will contribute to $A_{n>0,m}\left(
\omega \right) $, describing the creation of an electron in the system.
Substituting the expressions (\ref{GretT}) and (\ref{Kmt}) in (\ref{Anm}),
we get
\begin{eqnarray*}
iG_{n>0,m}^{Ret}\left( t\right) &=&\theta \left( t\right) \frac{1}{2\pi }%
\int_{0}^{2\pi }d\theta e^{i\left( n-1\right) \theta }e^{-i\mu ^{+}t} \\
&&\times e^{\sum_{\bar{n}=1}^{\infty }\frac{e^{-i\bar{n}\theta }}{\bar{n}}%
\int_{0}^{\infty }d\bar{q}\Phi _{m}^{2}\left( \bar{q}\right) e^{-iE_{\bar{n}%
}\left( \bar{q}\right) t}}.
\end{eqnarray*}
Now, the exponential function of the sum $\sum_{\bar{n}=1}^{\infty }$ can be
expanded in a series that, after the angular integration over $\theta $,
will result in a finite number of terms for each $n$. It is easy to see that
\begin{eqnarray}
iG_{n=1,m}^{Ret}\left( t\right) &=&\theta \left( t\right) e^{-i\mu ^{+}t},
\label{G0t} \\
iG_{n=2,m}^{Ret}\left( t\right) &=&\theta \left( t\right) e^{-i\mu
^{+}t}\int_{0}^{\infty }d\bar{q}\Phi _{m}^{2}\left( \bar{q}\right)
e^{-iE_{1}\left( \bar{q}\right) t},  \label{G3t2} \\
iG_{n=3,m}^{Ret}\left( t\right) &=&\frac{1}{2}\theta \left( t\right)
e^{-i\mu ^{+}t}\int_{0}^{\infty }d\bar{q}\Phi _{m}^{2}\left( \bar{q}\right) %
\left[ e^{-iE_{2}\left( \bar{q}\right) t}\right.  \notag \\
&&\left. +\int_{0}^{\infty }d\bar{q}^{\prime }\Phi _{m}^{2}\left( \bar{q}%
^{\prime }\right) e^{-i\left[ E_{1}\left( \bar{q}\right) +E_{1}\left( \bar{q}%
^{\prime }\right) \right] t}\right]  \notag
\end{eqnarray}
and so on for $n>3$. Therefore, we see that for each $iG_{n,m}^{Ret}\left(
t\right) $ we only have the contribution of the normal modes presenting $%
\bar{n}<n$. This behavior is general, occurring in any bosonized system that
presents a discrete levels structure. Recently, it has also been observed in
the treatment of the Luttinger liquid in a finite one-dimensional wire \cite
{Fab}, where the discrete levels structure came from the box-like boundary
conditions. We can also see that we do not have the contribution of the mode
$\bar{n}=n$. Therefore, according to (\ref{Anm}) and (\ref{G0t}), the
spectral function associated with the creation of a particle in the first
excited state, $n=1$, is given simply by
\begin{equation}
A_{n=1,m}\left( \omega \right) =\delta \left( \omega -\mu ^{+}\right) .
\label{Anmw1}
\end{equation}
In this situation the only energy level excited by the introduced particle
is the one given by the chemical potential, without any contribution from
the excitation of bosonic modes. This was expected as a consequence of the
first order approximation. Consider the inclusion of a particle in a state $%
\left| n=1,m\right\rangle ,$ as presented in Fig.(\ref{FigAw1}.a).
Again, since our ground state has no particle-hole pairs, there is
no empty state to which this particle could relax. Therefore, it
cannot excite (or decay into) any of the bosonic modes of the
system. The particle can neither move between the others channels,
once the number of particles in each channel must be conserved.
These means that the particle can not
``see'' the degeneracy breaking in $m$, associated with the dispersion $%
\Delta E_{n=1}\left( \mathbf{q}\right) $ in the level it is
included. It is easy to see that this is as a consequence of the
first order perturbation theory. Due to the angular momentum
conservation, the particle could jump to a channel $m^{\prime
}\neq m$, e.g. $m^{\prime }=m+1$, only exciting another particle
in another level $n$ above it, as presented in Fig.(\ref{Jump}).
However, this should be a virtual transition, once it does not
conserve energy.

According (\ref{Anm}) and (\ref{G3t2}), the other spectral functions will be
given by:
\begin{eqnarray}
A_{n=2,m}\left( \omega \right) &=&\int_{0}^{\infty }d\bar{q}\Phi
_{m}^{2}\left( \bar{q}\right) \delta \left( \omega -E_{1}\left( \bar{q}%
\right) -\mu ^{+}\right) ,  \notag \\
A_{n=3,m}\left( \omega \right) &=&\frac{1}{2}\int_{0}^{\infty }d\bar{q}\Phi
_{m}^{2}\left( \bar{q}\right) \left\{ \left[ \delta \left( \omega
-E_{2}\left( \bar{q}\right) -\mu ^{+}\right) +\right. \right.  \label{Anmw2}
\\
&&\left. \left. \int_{0}^{\infty }d\bar{q}^{\prime }\Phi _{m}^{2}\left( \bar{%
q}^{\prime }\right) \delta \left( \omega -E_{1}\left( \bar{q}\right)
-E_{1}\left( \bar{q}^{\prime }\right) -\mu ^{+}\right) \right] \right\}
\notag
\end{eqnarray}
and so on. It is clear, from the scheme of Fig.(\ref{FigAw1}), that the
above expressions correspond to the distribution (or decay) of a
single-particle excitation in a set of collective bosonic excitations of
total energy close to the free energy of the particle introduced. As the
energy of the introduced particle increases it excites a larger number of
collective modes. This kind of distribution is the essence of the
bosonization methods \cite{Voit2}. In the next section, after we include the
particle spin, we will present the graphic representation of some of this
spectral functions (see Fig.(\ref{FigAws})).

Analyzing the above expansion of $G_{n,m}^{Ret}\left( t\right) $ and
recalling that $\int_{0}^{\infty }d\bar{q}\Phi _{m}^{2}\left( \bar{q}\right)
=1$ we see that $A_{n,m}\left( \omega \right) $ obeys the sum rule $%
\int_{-\infty }^{\infty }d\omega A_{n,m}\left( \omega \right) =1$, for all $%
\left\{ n,m\right\} $.

The spectral density associated with the inclusion of a particle on the
position $\mathbf{r}$ is given by
\begin{equation}
A\left( \mathbf{r,}\omega \right) =-\frac{1}{\pi }ImG^{Ret}\left( \mathbf{r,}%
\omega \right) =\sum_{n=0}^{\infty }\sum_{m=0}^{N_{\phi }-1}\left| \langle
\mathbf{r}|n,m\rangle \right| ^{2}A_{n-\left( \nu -1\right) ,m}\left( \omega
\right) ,  \label{Arw1}
\end{equation}
where we use the relation (\ref{Gretr}) between $G^{Ret}\left( \mathbf{r}%
,t\right) $ and $G_{n-\left( \nu -1\right) ,m}^{Ret}\left( t\right) $. We
verify that $\int_{-\infty }^{\infty }d\omega A\left( \mathbf{r,}\omega
\right) =1$, once we have the correct normalization $\sum_{n,m}\left|
\langle \mathbf{r}|n,m\rangle \right| ^{2}=1$. Particularly, for the
creation of a particle at the origin, $\mathbf{r}=\mathbf{0}$, we have
\begin{equation}
A\left( \mathbf{r}=\mathbf{0,}\omega \right) =\frac{1}{\left( 2\pi l\right)
^{2}}\sum_{n=0}^{\infty }A_{n-\left( \nu -1\right) ,n}\left( \omega \right) ,
\label{ParO}
\end{equation}
because $\langle \mathbf{0}|n,m\rangle =\delta _{n,m}/\sqrt{2\pi l^{2}}$.
One particle created at $\mathbf{r=0}$ $\,$add to the system an angular
momentum $\Delta L_{z}=0$. Therefore, it must decay only into the modes with
$n=m$, which also present zero angular momentum and are indeed the only
terms left in (\ref{Arw1}). We will see the graphic representation of $%
A\left( \mathbf{r=0,}\omega \right) $ in Fig.(\ref{FigAwr0}), after the
inclusion of the spin degree of freedom.

\subsection{Including the Electron Spin}

The details of the introduction of the electron spin, $s=\pm 1$, are
presented in the Appendix D. For the Green's function we get
\begin{eqnarray}
iG_{n,m,s}^{Ret}\left( t\right) &=&\theta \left( t\right) \int_{0}^{2\pi
}d\theta e^{in\theta }\left[ e^{-i\mu _{s}^{+}t}e^{-i\theta }M_{m,\theta
}\left( t\right) \right.  \notag \\
&&\left. +e^{-i\mu _{s}^{-}t}M_{m,\theta }^{\ast }\left( t\right) \right] ,
\label{Gretnms} \\
M_{m,\theta }\left( t\right) &=&\frac{1}{2\pi }e^{\frac{1}{2}%
\sum_{n=1}^{\infty }\frac{1}{n}e^{-in\theta }\int_{0}^{\infty }d\bar{q}\Phi
_{m}^{2}\left( \bar{q}\right) \left( e^{-iE_{n}^{\left( \rho \right) }\left(
\bar{q}\right) t}+e^{-iE_{n}^{\left( \sigma \right) }\left( \bar{q}\right)
t}\right) },  \notag
\end{eqnarray}
with $E_{n}^{\left( i\right) }$ given by (\ref{HTS}).

\begin{figure}[tbp]
 \includegraphics[width=0.65%
\columnwidth,keepaspectratio]{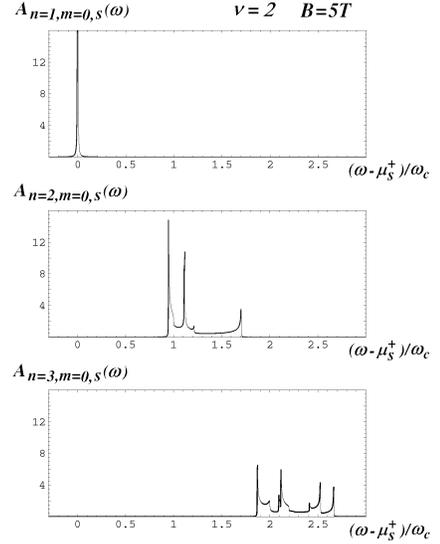}
\par
 \caption{Examples of spectral functions $A_{n,m=0,s}\left( \protect\omega
\right) $ for a system with filling factor $\protect\nu =2$ and a typical
transverse magnetic field of $5T$ in a GaAs heterojunction, where we have $%
\hbar \protect\omega _{c}\left[ K\right] \simeq 20B\left[ T\right] $ and $%
e^{2}/\protect\epsilon l$ $\left[ K\right] \simeq 50\protect\sqrt{B\left[ T%
\right] }$. The $\protect\delta $-functions in (\ref{subsf2}) where
substituted by a Lorentzians of width $10^{-3}\protect\omega _{c}$.}
\label{FigAws}
\end{figure}

The spectral function $A_{n,m,s}\left( \omega \right) $ presents the same
structure of the equations (\ref{Anmw1}-\ref{Anmw2}) under the following
substitutions:
\begin{widetext}
\begin{eqnarray}
\delta \left( \omega -E_{n}\left( \bar{q}\right) -\mu ^{+}\right)
&\rightarrow &\frac{1}{2}\sum_{i=\rho ,\sigma }\delta \left( \omega
-E_{n}^{\left( i\right) }\left( \bar{q}\right) -\mu _{s}^{+}\right) ,
\label{subsf2} \\
\delta \left( \omega -E_{n}\left( \bar{q}\right) -E_{n}\left( \bar{q}%
^{\prime }\right) -\mu ^{+}\right) &\rightarrow &\frac{1}{4}\left[
\sum_{i=\rho ,\sigma }\delta \left( \omega -E_{n}^{\left( i\right) }\left(
\bar{q}\right) -E_{n}^{\left( i\right) }\left( \bar{q}^{\prime }\right) -\mu
_{s}^{+}\right) +2\delta \left( \omega -E_{n}^{\left( \rho \right) }\left(
\bar{q}\right) -E_{n}^{\left( \sigma \right) }\left( \bar{q}^{\prime
}\right) -\mu _{s}^{+}\right) \right] ,  \notag
\end{eqnarray}
and so on. The functions $A_{n,m,s}\left( \omega \right) $ also obey the sum
rule $\int_{-\infty }^{\infty }d\omega A_{n,m,s}\left( \omega \right) =1$,
for all $\left\{ n,m,s\right\} $.

In Fig.(\ref{FigAws}) we present examples of the form acquired by these
spectral functions. The spectral function $A_{n=1,m,s}\left( \omega \right) $
corresponds simply to a $\delta -$function centered at the chemical
potential energy $\mu _{s}^{+}$. From $A_{n=2,m,s}\left( \omega \right) $ on
to higher values of $n$ we start to have a complex energy distribution. The
particle excites simultaneously, according the scheme of Fig.(\ref{FigAw1}%
.b), the charge and spin modes, associated with $E_{n=1}^{\left( \rho
\right) }\left( \bar{q}\right) $ and $E_{n=1}^{\left( \sigma \right) }\left(
\bar{q}\right) $. The spectral function $A_{n=2,m,s}\left( \omega \right) $
(see (\ref{Anmw2})) can also be written as:

\begin{equation}
A_{n=2,m,s}\left( \omega \right) =\frac{1}{2}\sum_{i=\rho ,\sigma
}\int_{0}^{\infty }\frac{dE\left( \bar{q}\right) }{\left| \frac{d}{d\bar{q}}%
E\left( \bar{q}\right) \right| }\Phi _{m}^{2}\left( \bar{q}\right) \delta
\left( \omega -E_{n=1}^{\left( i\right) }\left( \bar{q}\right) -\mu
_{s}^{+}\right) \equiv \frac{1}{2}\sum_{i=\rho ,\sigma }\sum_{\bar{q}%
\leftrightarrow E_{n=1}^{\left( i\right) }\left( \bar{q}\right) +\mu
_{s}^{+}=\omega }\frac{\Phi _{m}^{2}\left( \bar{q}\right) }{\left| \frac{d}{d%
\bar{q}}E_{n=1}^{\left( i\right) }\left( \bar{q}\right) \right| },
\label{AdE}
\end{equation}
\end{widetext}
where we have performed the integration $\int_{0}^{\infty }d\bar{q}=$ $%
\int_{0}^{\infty }\frac{dE\left( \bar{q}\right) }{\left| \frac{d}{d\bar{q}}%
E\left( \bar{q}\right) \right| }$ and
$\sum_{\bar{q}\leftrightarrow E_{n=1}^{\left( i\right) }\left(
\bar{q}\right) +\mu _{s}^{+}=\omega }$ means a infinite sum over
all the $\bar{q}$ values such that $E_{n=1}^{\left(
i\right) }\left( \bar{q}\right) +\mu _{s}^{+}$ corresponds to the given $%
\omega $. So $A_{n=2,m,s}\left( \omega \right) $ will present peaks
(integrable divergences), with weights proportional to $\Phi _{m}^{2}\left(
\bar{q}\right) /2$ where $dE_{n=1}^{\left( i\right) }\left( \bar{q}\right) /d%
\bar{q}=0$ and the system presents maxima in its density of states. This
behavior is similar to the behavior of the dynamical structure factor, $%
S\left( q,\omega \right) $, measured by inelastic light-scattering
\cite {PinkKal,Eric} in integer quantum Hall system, which confirm
the existence of the rotons´ minima in the dispersion of the
collective modes. We can easily calculate $S\left( q,\omega
\right) $ within the bosonization procedure, but since we obtain
the same results presented in \cite{PinkKal} we do not present the
calculation here.

\begin{figure}[tbp]
 \includegraphics[width=1.0%
\columnwidth,keepaspectratio]{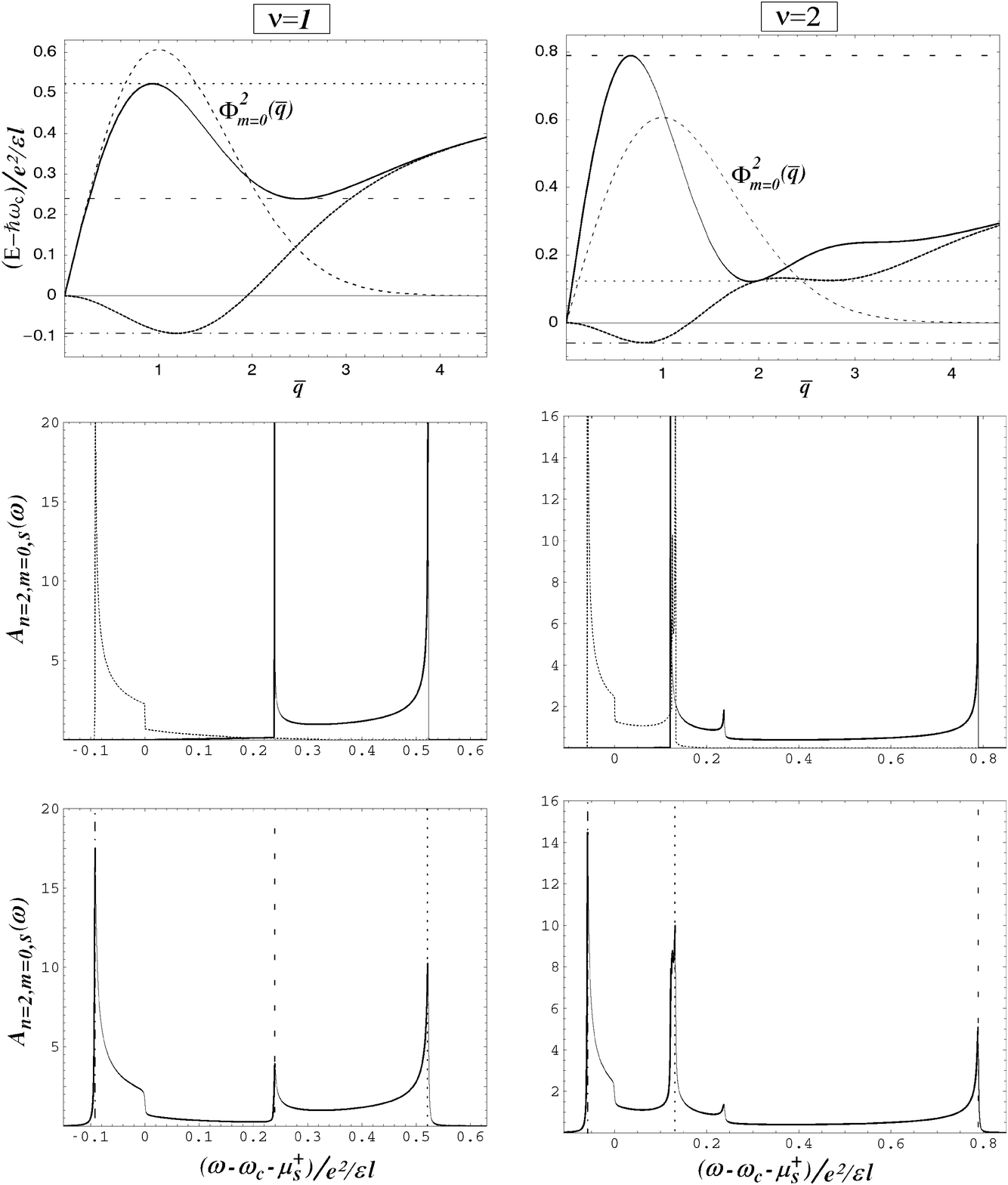}
\par
 \caption{Spectral functions $A_{n=2,m=0,s}\left( \protect\omega \right) $
for different filling factors $\protect\nu $. In the second line the $%
\protect\delta $-functions in (\ref{subsf2}) where substituted by
Lorentzians of width $10^{-6}\protect\omega _{c}$, stressing the
integrable divergence at the local extremes of the dispersions
$E_{n=1}^{\left( i\right) }\left( \bar{q}\right) $. The charge
(continuous) and spin (dotted) contributions are represented
separately. The third line represents the
addition of the two previous contributions, but with Lorentzians of width $%
10^{-3}\protect\omega _{c}$, such that we can see the relative weight of the
different peaks. Horizontal and vertical lines with the same dotted pattern
correspond to the same energy ($\protect\omega =E$ with $\hbar =1$).}
\label{FigAwSC}
\end{figure}

The expression (\ref{AdE}) is useful to understand the behavior of $%
A_{n,m,s}\left( \omega \right) $. However, the more direct way to
evaluate it is substituting $\delta \left( \omega -E_{n=1}^{\left(
i\right) }\left( \bar{q}\right) -\mu _{s}^{+}\right) $ by a
Lorentzian of negligible width. This procedure allows us to
compare the weight of the different peaks, while keeping the area
under each one of them constant. In Fig.(\ref{FigAwSC}) we present
the behavior of $A_{n=2,m=0,s}\left( \omega \right) $ for two
different filling factors $\nu $. We see peaks in the energies of
the extreme points of the charge and spin fluctuation modes and
the energy of the particle being distributed between the minimum
of $E_{n=1}^{\left( \sigma \right) }\left( \bar{q}\right) $ and
the maximum of $E_{n=1}^{\left( \rho \right) }\left(
\bar{q}\right) $. This behavior is analogous to the distribution
of energy of an added particle in the Tomonaga-Luttinger model \cite{Voit2}%
. There the energy of a particle with momentum $q$ is partially distributed
between the charge and spin modes, presenting dispersions $v_{\rho }q$ and $%
v_{\sigma }q$, and also over energies above and below these ones, due to the
existence of the anomalous dimension. Here we do not have any effect of an
anomalous dimension, see sec.(\ref{Ground}), however, both dispersion
relations, $E_{n}^{\left( \rho \right) }\left( \bar{q}\right) $ and $%
E_{n}^{\left( \sigma \right) }\left( \bar{q}\right) $, present a
non-monotonic behavior, giving rise to the many peaks in the spectral
function. We also have the presence of the function $\Phi _{m}^{2}\left(
\bar{q}\right) $, which distributes the energy according to the relation
between an excitation in the guiding center $m$ and one with wave vector $%
\bar{q}$, which we will discuss in sec.\ref{F}.

\subsection{Non-normal System}

In a non-interacting system, for $n\geq 1$, we would have $\left( \hbar
=1\right) $%
\begin{equation*}
A_{n,m,s}^{0}\left( \omega \right) =\delta \left( \omega -\left( n-1\right)
\omega _{c}-\mu _{0,s}^{+}\right) ,
\end{equation*}
where $\mu _{0,s}^{+}=\hbar \omega _{c}+s\Delta E_{z}$, $\Delta E_{z}$ is
the Zeeman energy, and the retarded Green's function would be given by
\begin{equation}
G_{n,m,s}^{0,Ret}\left( \omega \right) =\frac{1}{\omega -\left( n-1\right)
\omega _{c}-\mu _{0,s}^{+}+i\eta }.  \label{G0ret}
\end{equation}
Now let us study what happens in the interacting system. For $n=1$, the
introduction of the interaction results in a simple renormalization of the
chemical potential $\mu _{0,s}^{+}$ to $\mu _{s}^{+}$, similar to (\ref
{PotQ2}). For $n>1$ we have the spectral functions with the (integrable)
divergences of Figs.(\ref{FigAws}) and (\ref{FigAwSC}). If we take the
Fourier transform of $G_{n=2,m,s}^{Ret}\left( t\right) $, given by (\ref
{Gretnms}), we get
\begin{equation*}
G_{n=2,m,s}^{Ret}\left( \omega \right) =\frac{1}{2}\sum_{i=\rho ,\sigma
}\int_{0}^{\infty }d\bar{q}\frac{\Phi _{m}^{2}\left( \bar{q}\right) }{\omega
-E_{n=1}^{\left( i\right) }\left( \bar{q}\right) -\mu _{s}^{+}+i\eta }.
\end{equation*}
So, instead of a single pole, as in (\ref{G0ret}), we have a
continuum of poles which give rise to some integrable divergences
in the spectral functions. Therefore, we see that the effect of
the interaction on the non-interacting spectral function is very
different from shifting and broadening the non-interacting peak.
It generates peaks with different energies and weights, associated
with both charge and spin fluctuations. In
normal systems \cite{Noz,Mahan}, for each non-interacting eigenstate, $%
\left| n,m,s\right\rangle $, we would obtain a single solution of the
equation
\begin{equation*}
\omega -\left( n-1\right) \omega _{c}-\mu _{s}^{+}-Re\Sigma _{ret}\left(
\left| n,m,s\right\rangle \mathbf{,}\omega -\mu _{s}^{+}\right) =0,
\end{equation*}
and this would be the energy of the quasi-particle in the interacting
system. Here $\Sigma _{ret}$ is the retarded self-energy \cite{Noz,Mahan}.
However, in our case the above equation presents many solutions, one for
each divergent peak in $A_{n,m,s}\left( \omega \right) $. So, in general, we
see that we have a non-normal system. However, in our treatment, there are
two exceptions. The first is the case associated with the inclusion of a
particle in the lower allowed energy, when $A_{n=1,m,s}\left( \omega \right)
$ is given by a simple $\delta $-function. The second is the case $%
A_{n,m\rightarrow N_{\Phi }\gg 1,s}\left( \omega \right) $, that we will see
in the next section.

\subsection{Influence of the Guiding Center Position\label{F}}

\begin{figure}[tbp]
\includegraphics[width=1.0%
\columnwidth,keepaspectratio]{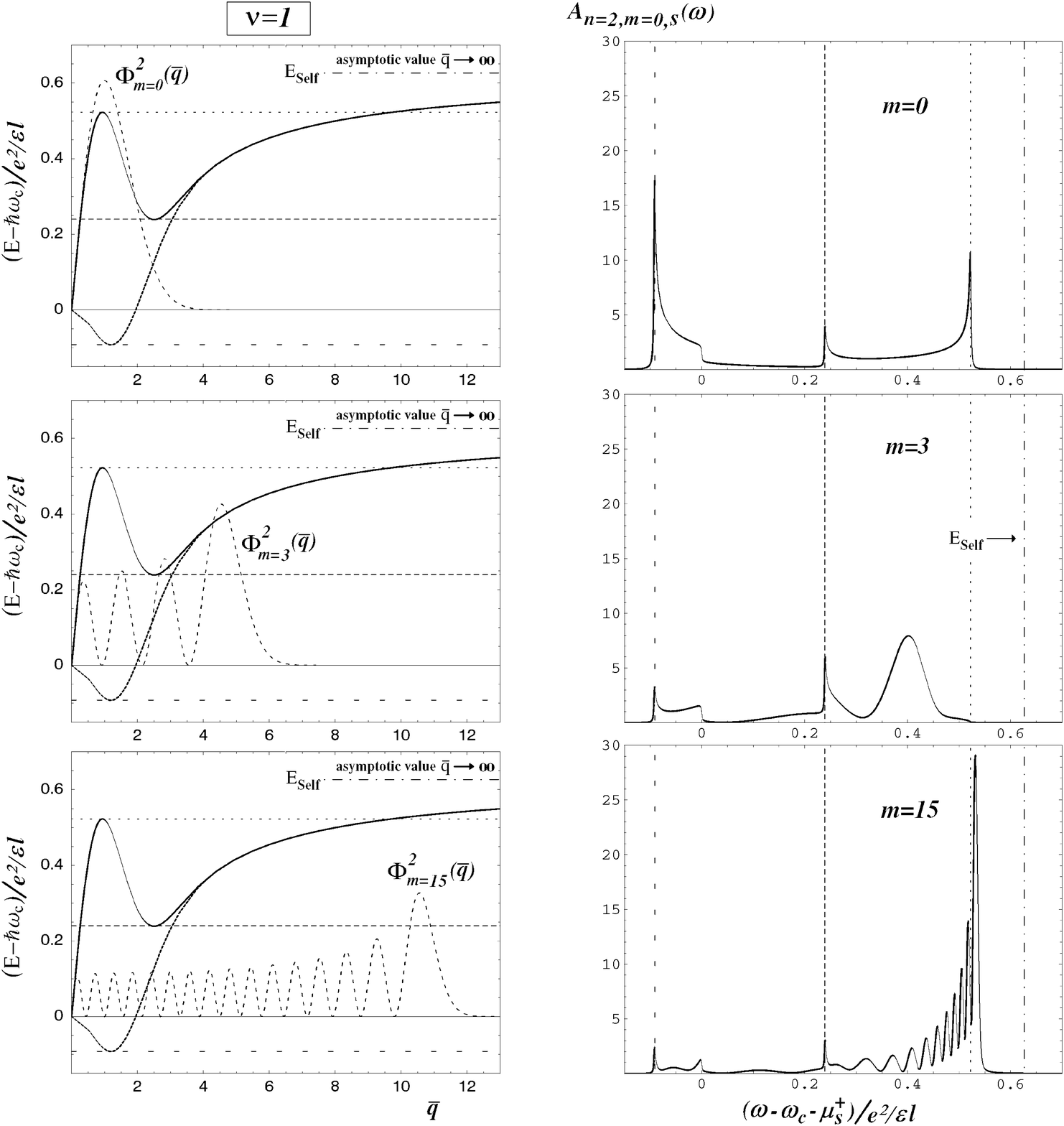}
\par
 \caption{Spectral function $A_{n=2,m}\left( \protect\omega \right) $ for $%
m=0,3$ and $15,$ in a system presenting a filling factor $\protect\nu =1$.
Horizontal and vertical lines with the same dotted pattern correspond to the
same energy. The vertical line indicated as $E_{Self}$ would be the position
of the peak in the limit $m\rightarrow N_{\Phi }\gg 1$.}
\label{Fmvar}
\end{figure}

\begin{figure}[ptb]
 \includegraphics[width=1.0%
\columnwidth,keepaspectratio]{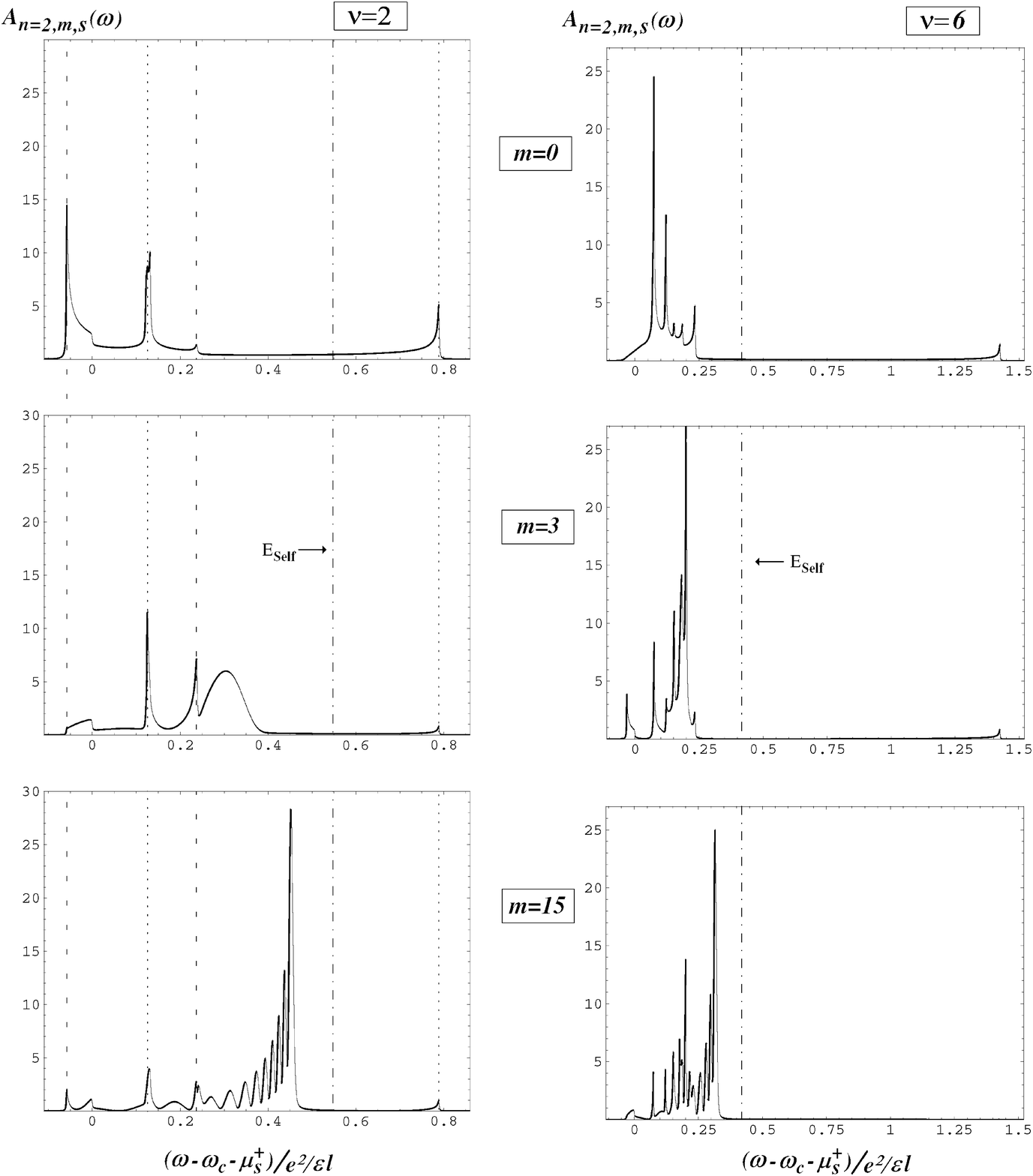}
\par
 \caption{Spectral functions $A_{n=2,m,s}\left( \protect\omega\right) $ for
different $m $ values in a system presenting filling factors $\protect\nu=2$
and $\protect\nu=6$. The vertical dotted lines correspond to local extremes
of the normal modes, as in the figure before, except for the vertical line
indicated as $E_{Self}$, corresponding to the asymptotic value.}
\label{FigAwnuvar}
\end{figure}

So far, in analyzing the spectral functions we have concentrated
on the influence of the varying densities of states. However, we
have also mentioned the influence of the guiding center position,
defined by $m$. The function $\Phi _{m}^{2}\left( \bar{q}\right) $
weights the importance of the density of
states at each point. The spectral functions presented in Figs.(\ref{FigAws}%
) and (\ref{FigAwSC}) were calculated with $m=0$. In this case $\Phi
_{m=0}^{2}\left( \bar{q}\right) $ does not present oscillations and limits
the influence of the densities of states to the range of medium (not too
large) $\bar{q}^{^{\prime }}s$. However, once we increase the value of $m$,
the function $\Phi _{m}^{2}\left( \bar{q}\right) $ spreads and starts to
present $m+1$ oscillations, with the last maximum being the largest one, see
Fig.(\ref{Fmvar}). Once the dispersion relations became softer at large $%
\bar{q}$'s this region starts to concentrate most part of the spectral
weight. Consequently, as shown in Fig.(\ref{Fmvar}), the spectral function $%
A_{n=2,m,s}\left( \omega \right) $ starts to present weight at
larger frequencies, with a peak much more pronounced that the one
in $\Phi _{m}^{2}\left( \bar{q}\right) $. In the limit
$m\rightarrow N_{\Phi }\gg 1$ the peak will be concentrated at the
asymptotic value $\omega _{c}+\mu _{s}^{+}+E_{Self}^{n=2}$, with a
characteristic modulation. It is clear that, in this limit, more
general spectral functions, $A_{n,m,s}\left( \omega \right) $,
would also present themselves concentrated at well defined
energies (shifted from the non-interacting values) and we will not
see any effect of the decomposition into charge and spin
fluctuations, once both present the same asymptotic limit. Systems
presenting higher filling factors, as in Fig.(\ref{FigAwnuvar}),
present a larger number of peaks, because of the $\nu $
oscillations in $E_{n}^{\left( \rho \right) }\left( \bar{q}\right)
$ and $E_{n}^{\left( \sigma \right) }\left( \bar{q}\right) $, but
their dispersions relations still  converge to their asymptotic
values in the limit $m\rightarrow N_{\Phi }\gg 1$.

We believe that this result can be understood as follows. A
particle in a state $\left| n,m\right\rangle $ presents a probability
density of width $\sqrt{2n+1}l$, spread over a circle of radius $\sqrt{2m+1}%
l $ \cite{kivelson}. Therefore, for not too large values of $n$,
creating a particle in a state written in terms of $m$$\sim
N_{\Phi }$ means to create a particle in a large radius, close to
the edge of the system. Most of the electron-hole pairs excited by
the introduction of this particle should also spread over large
radia, where their net wave function overlap is small and they do
not fell much the presence of each other. Therefore, the energies
excited in the system will be associated with electrons and holes
far apart, $E_{Self}^{n}$. However, the particles created close to
the system's origin, with states written in terms of small $m$'s,
would present a much more complex structure in their spectral
functions, because the electron-hole pairs excited will present a
much larger correlation. Below, we will explicitly calculate the
spectral function of a particle added at the system´s origin.

\begin{figure}[tbp]
 \includegraphics[width=0.7%
\columnwidth,keepaspectratio]{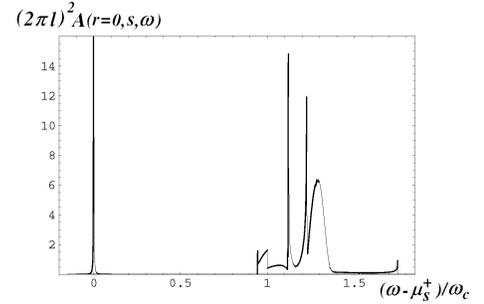}
\par
 \caption{Spectral function for a particle created at the center of a GaAs
heterojunction presenting filling factor $\protect\nu =2$ under a transverse
magnetic field $B=7T$.}
\label{FigAwr0}
\end{figure}

The inclusion of the spin in $G^{Ret}\left( \mathbf{r},t\right) $ is made by
substituting $G_{n,m}^{Ret}\left( t\right) $ by $G_{n,m,s}^{Ret}\left(
t\right) $ in (\ref{Gretr}):
\begin{equation*}
G^{Ret}\left( \mathbf{r},t,s\right) =-i\sum_{n,m}\left| \langle \mathbf{r}%
|n,m\rangle \right| ^{2}G_{n-\left( \nu -1\right) ,m,s}^{Ret}\left( t\right)
.
\end{equation*}
The same must be done for the calculation of $A\left(
\mathbf{r},s,\omega \right) $. For a particle added at the
system´s origin we have the expression equivalent to (\ref{ParO}):
\begin{equation}
A\left( \mathbf{r=0,}s,\omega \right) =\frac{1}{\left( 2\pi l\right) ^{2}}%
\sum_{n=0}A_{n-\left( \nu -1\right) ,n,s}\left( \omega \right) .  \label{Ar0}
\end{equation}
Therefore, $A\left( \mathbf{r=0,}s,\omega \right) $ will present many peaks.
This is similar to the sum of the different peaks presented in Fig.(\ref{FigAws}%
), but with each peak being modulated by the corresponding $\Phi
_{m=n}^{2}\left( \bar{q}\right) $ instead of the same $\Phi
_{m=0}^{2}\left( \bar{q}\right) $. The terms presenting $%
n\leq \nu -1$ will generate contributions to frequencies $\omega <0$,
describing the effect of creating a hole in the system. They are
similar to the ones in the region $\omega >0$. Fig.(\ref{FigAwr0}) presents
the first two peaks of $A\left( \mathbf{r=0,}s,\omega \right) $ in the
region $\omega >0$ as an example.

Expression (\ref{Ar0}) also indicates that as the filling factor increases
the spectral function develops better defined peaks in the $\omega >0$
region. In a system with $\nu \gg 1$ the spectral function $A\left(
\mathbf{r=0,}s,\omega \right) $ at frequencies $\omega \gtrsim \omega _{c}$
(associated with $\left| n\right| \gtrsim \nu $) will be governed by the
functions $\Phi _{m=n\gtrsim \nu \gg 1}^{2}\left( \bar{q}\right) $. So the
system will tend to present a set of well defined peaks, shifted from the
non-interacting values, as in a normal system.

\section{Conclusions}

In this work we presented a model and a method for the study of the integer
quantum Hall systems presenting both spin sublevels filled $\left( {\bf S}=%
{\bf 0}\right) $. We showed that, to first order in $\gamma =\left(
e^{2}/\varepsilon l\right) /$ $\hbar \omega _{c}$, our model of interacting
one-dimensional channels within the {\it forward scattering approximation,}
which keeps constant the net number of particles in each channel, is
equivalent to the {\it strong field approximation}. We saw how this model
can be solved, with some well controlled approximations, through an
appropriate variation of the {\it Landau level bosonization method}. Using
this set of model and method we determined the collective excitations of the
system: spin and charge fluctuations, dependent on the direct and exchange
interactions. We obtained the same dispersion relations as those presented
by Kallin and Halperin \cite{Halp} and MacDonald \cite{MacDH} for the
excitations close to the cyclotron frequency $\omega _{c}$.

From these non-monotonic dispersion relations and variable density of
states, we determine the single particle spectral functions through the
bosonic {\it Mattis-Mandelstam} representation of the fermionic operator.
For states created close to the central region of the system, these
functions presented the characteristic behavior of a strongly correlated
{\it non-normal system}, including a spin and charge spectral decomposition
and absence of quasi-particle excitations that keep similarities with the
ones of a non-interacting system. However, for states close to the edge of
the system these functions tend to present single peaks, shifted from the
non-interacting values and with a characteristic envelope, like in a {\it %
normal system}. It is clear that, in our treatment, we do not consider the
effects of the confining potential in the Hall system, which generate
gapless edge excitations \cite{HalpOri} and are fundamental to a complete
description of the edge states . However, independently of the additional
effects of the confining potential, what we see is that the bulk
quasi-particle states seem to be much more strongly coupled to the
collective excitations in the system than the edge ones, which seems to
behave like normal Fermi liquid particles, even in higher energies.

The experimental verification of the results we predicted seems to require a
control over the state of the created particles higher than the one
available in current experiments. However, if as we predicted here, the
spectral functions for states created at the central region and those close
to edge of the system have such a distinct behavior, this should not be too
difficult to observe.

As we mentioned in the introduction and in Sec.(III.A), one of the
advantages of the present model for the quantum Hall systems is that it
presents a prescription for its improvement, once there are methods
developed to treat the net particle transfer between neighbor channels in a
bosonic description \cite{RefI}. We saw that the previous perturbative
methods for IQH\ systems \cite{Halp,MacDH} used $\gamma $ as the expansion
parameter. However, they do not give a systematic description, order by order,
in this parameter. Indeed, not even the first order in $\gamma $ is fully
described, indicating that this may not be the better expansion parameter.
So, we propose, start from the forward scattering approximation, which is
already a good initial description as we saw, and try to get higher order
corrections taking the net particle transfer between ``$m$'' neighbor
channels as the expansion parameter. Once this transfer would be between
channels with non-zero relative angular momentum, we suspect that the
effective transfer potential can be related with some kind of {\it Haldane
pseudo-potential} \cite{Hald5}, which, at least in the fractional quantum
Hall systems, are known to be the really effective expansion parameters.
Certainly, with the inclusion of the transfer between neighbor channels, we
will have a Hamiltonian whose ground state will present particle-holes
pairs, taking into account the Landau level mixing, which would be very
convenient in many applications. In this case, the spectral functions $%
A_{n=1}\left( \omega \right) $ should already present a complex structure,
in contrast to the $\delta $-function that we have got.

We also mentioned the fundamental importance of the collective modes in the
FQHE, evidenced by the Hamiltonian theory of Shankar and Murthy \cite{ShaN},
which have many similarities with a bosonization method. Their fermionic
field associated with the composite particle, $\psi _{CP}$, is very similar
to a {\it Mattis-Mandelstam} representation associated with the
magnetoplasmon excitation. This seems to us another evidence that systematic
bosonization methods created for quantum Hall system should be explored.

Finally, we note that the one dimensional characteristics that seem to be
retained in the 2DEG under a strong perpendicular magnetic field, the
influence of the collective modes in the quantum Hall quasi-particle
properties and the possibility of using the relative angular momentum
between particles as an expansion parameter in a field theory, probably can
still be further explored in quantum Hall systems. Here we presented a
method that partially explores the first two characteristics, indicates a
way that can be followed in the investigation of the third one and can be improved for further applications.

\begin{acknowledgments}

MRC and AOC acknowledge Funda\c{c}\~ao de Amparo \`a Pesquisa do
Estado de S\~ao Paulo (FAPESP) for the financial support. MRC also
acknowledge Financiadora de Estudos e Projetos (FINEP) and
Minist\'erio da Ci\^encia e Tecnologia (MCT). HW and AOC also
acknowledge Conselho Nacional de Desenvolvimento Cient\'ifico e
Tecnol\'ogico (CNPq) for partial support.

\end{acknowledgments}

\appendix
\section{Some Properties of the Functions $G_{n,m^{\prime}}\left(\mathbf{q%
}\right)$}

In this Appendix we just list some relations among the $G_{n,m}\left(
\mathbf{q}\right) $ functions \cite{allan}, with $\mathbf{q}=q_{x}+iq_{y}$,
which are useful for the matrix element manipulation.

Hermitian Conjugate:
\begin{equation}
G_{m^{\prime },m}\left( \mathbf{q}\right) =G_{m,m^{\prime }}^{\ast }\left( -%
\mathbf{q}\right) =G_{m,m^{\prime }}\left( \mathbf{q}^{\ast }\right) .
\label{GH}
\end{equation}

Matrix Product:
\begin{equation}
\sum_{l}G_{m^{\prime },l}(\mathbf{q})G_{l,m}(\mathbf{q}^{\prime })=e^{-\frac{%
\mathbf{q}^{\ast }\mathbf{q}^{\prime }}{2}}G_{m^{\prime },m}(\mathbf{q}+%
\mathbf{q}^{\prime }).  \label{Soma}
\end{equation}

Full Landau Level Rule:
\begin{equation}
\sum_{m=0}^{\infty }G_{m.m}\left( \mathbf{q}\right) =N_{\Phi }\delta _{%
\mathbf{q,0}}=\frac{\left( 2\pi \right) ^{2}}{S}N_{\Phi }\delta \left(
\mathbf{q}\right) .  \label{Fll}
\end{equation}

Orthogonality:
\begin{equation*}
\int d^{2}\mathbf{q}e^{-q^{2}/2}G_{m.m^{\prime }}\left( \mathbf{q}\right)
G_{n,n^{\prime }}\left( \mathbf{q}^{\ast }\right) =\frac{1}{2\pi }\delta
_{m,n}\delta _{m^{\prime },n^{\prime }}.
\end{equation*}

Matrix Elements:

\begin{equation}
\left\langle n,m\right| e^{-i\mathbf{q\cdot r}}\left| n^{\prime },m^{\prime
}\right\rangle =e^{-\frac{\left| l\mathbf{q}\right| ^{2}}{2}}G_{n,n^{\prime
}}\left( l\mathbf{q}^{\ast }\right) G_{m,m^{\prime }}\left( l\mathbf{q}%
\right) .  \label{Fou}
\end{equation}

\section{The interaction matrix elements}

To obtain the matrix elements of the interaction Hamiltonian in the forward
scattering approximation, eq.(\ref{Htot}), one has to pair the creation and
annihilation fermionic operators of (\ref{HI}) in all possible ways and set
the same guiding center $m$ for both. This is easily accomplished by
defining a generalized density operator
\begin{equation}
\rho \left( \mathbf{q,x}\right) =e^{-\frac{\bar{q}^{2}}{2}}\sum_{\left\{
n,m\right\} }G_{n,n^{\prime }}\left( \mathbf{\bar{Q}}^{\ast }\right)
G_{m,m^{\prime }}\left( \mathbf{\bar{q}}\right) c_{n,m}^{\dagger
}c_{n^{\prime },m^{\prime }},  \label{rhoxi}
\end{equation}
with $\mathbf{\bar{q}}=\mathbf{q}l$ , $\mathbf{\bar{Q}=\bar{q}}+i\mathbf{%
\bar{x}}$ and $\mathbf{\bar{x}=x}/l$, and finding its bosonic representation
in terms of the operators $b_{n,m}$ or even directly in terms of $%
b_{n}\left( q\right) $, as we will do here. It is straightforward, though a
little cumbersome, to show that the two ways of pairing the creation and
annihilation operators of the Hamiltonian that yield the direct and exchange
terms as follows
\begin{equation}
H_{I}^{Dir}=\frac{1}{2}\frac{1}{\left( 2\pi \right) ^{2}}\int d^{2}q\,\tilde{%
V}\left( \mathbf{q}\right) \rho ^{\dagger }\left( \mathbf{q}\right) \rho
\left( \mathbf{q}\right) -\frac{1}{2}V(\mathbf{0})\hat{N}_{T},
\label{hdirro2}
\end{equation}
\begin{eqnarray}
H_{I}^{Ex} &=&-\frac{1}{2}\frac{1}{\left( 2\pi \right) ^{2}}\int d^{2}q\int
d^{2}xV\left( x\right) e^{-\frac{\bar{x}^{2}}{2}}e^{i(\mathbf{\bar{q}\cdot
\bar{x}}^{\ast }-\mathbf{\bar{q}}^{\ast }\mathbf{\cdot \bar{x}})/2}  \notag
\\
&&\times \rho \left( \mathbf{q,x}\right) \rho ^{\dagger }\left( \mathbf{q,x}%
\right) +\frac{1}{2S}\tilde{V}\left( 0\right) \hat{N}_{T},  \label{HIEX}
\end{eqnarray}
where $\rho \left( \mathbf{q}\right) =\rho \left( \mathbf{q,x=0}\right) $ is
the electron density operator and we introduced the total particle number
operator $\hat{N}_{T}=\sum_{n,m}c_{n,m}^{\dagger }c_{n,m}$. To obtain the
exchange term \cite{Hayko}, the matrix element in (\ref{HI}) must be
expressed in the position basis, $\left| \mathbf{r,r}^{\prime }\right\rangle
$, using (\ref{G}). Then, in the resulting expression, we replace the
integration variables, $\mathbf{r}$ and $\mathbf{r}^{\prime }$, by the
relative, $\mathbf{x}=\mathbf{r}-\mathbf{r}^{\prime }$, and center of mass, $%
\mathbf{X=}\left( \mathbf{r}+\mathbf{r}^{\prime }\right) /2$, coordinates
evaluate the integral with respect to the latter and use the relation (\ref
{Soma}) from Appendix A.

To obtain the bosonic representation of $\rho \left( \mathbf{q,x}\right) $
we first rewrite the sum over $\left\{ n,n^{\prime }\right\} $ as $%
\sum_{n,n^{\prime }}=\sum_{n>n^{\prime }}+\sum_{n<n^{\prime }}+(n=n^{\prime
})$, doing $n\equiv p+n$ and $n^{\prime }\equiv n$ in the terms where $%
n>n^{\prime }$ and $n \rightleftarrows n^{\prime }$  when
$n<n^{\prime }$. At the same time we can make the forward
scattering approximation, $m=m^{\prime }$, to get
\begin{eqnarray}
\rho \left( \mathbf{q,x}\right) &=&e^{-\frac{\bar{q}^{2}}{2}}\sum_{n>0,p}%
\left[ G_{p,p+n}\left( \mathbf{\bar{Q}}^{\ast }\right) \sum_{m}G_{m,m}\left(
\mathbf{\bar{q}}\right) \right. c_{p,m}^{\dagger }c_{p+n,m}  \notag \\
&&\left. +G_{p+n,p}\left( \mathbf{\bar{Q}}^{\ast }\right)
\sum_{m}G_{m,m}\left( \mathbf{\bar{q}}\right) c_{p+n,m}^{\dagger }c_{p,m}%
\right]  \label{nex2} \\
&&+\hat{\rho}_{N}\left( \mathbf{q,x}\right) ,  \notag
\end{eqnarray}
where
\begin{equation*}
\hat{\rho}_{N}\left( \mathbf{q,x}\right) =e^{-\frac{\bar{q}^{2}}{2}%
}\sum_{n,m}G_{n,n}\left( \mathbf{\bar{Q}}^{\ast }\right) G_{m,m}\left(
\mathbf{\bar{q}}\right) c_{n,m}^{\dagger }c_{n,m}\,.
\end{equation*}

Now we approximate $G_{p,p+n}\left( \mathbf{\bar{Q}}^{\ast }\right) \simeq
\left. G_{p,p+n}\left( \mathbf{\bar{Q}}^{\ast }\right) \right| _{p=\nu -1}$
since we are restricted to the low energy excitations close to the higher
filled Landau level, $\nu -1$. Using simultaneously (\ref{b+}) and (\ref{b1}%
), we get:
\begin{eqnarray}
\rho \left( \mathbf{q,x}\right) &=&\frac{e^{-\frac{\bar{q}^{2}}{4}}}{\sqrt{%
\bar{q}}}\sum_{n}\sqrt{n}\left[ G_{\nu -1,\nu -1+n}\left( \mathbf{\bar{Q}}%
^{\ast }\right) b_{n}\left( \bar{q}\right) \right.  \label{roexb1} \\
&&\left. +G_{\nu -1+n,\nu -1}\left( \mathbf{\bar{Q}}^{\ast }\right)
b_{n}^{\dagger }\left( \bar{q}\right) \right] +\hat{\rho}_{N}\left( \mathbf{%
q,x}\right) .  \notag
\end{eqnarray}

The bosonization of $\hat{\rho}_{N}\left( \mathbf{q,x}\right) $ is done in
the same way as the bosonization of $H_{0}$ \cite{VanD}, eq.(\ref{H0b}),
because both have the same form except for the expression $G_{n,n}\left(
\mathbf{\bar{Q}}\right) G_{m,m}\left( \mathbf{\bar{q}}\right) $ which
replaces the simpler function $n$. Calculating the commutator
\begin{widetext}
\begin{eqnarray*}
\left[ \hat{\rho}_{N}\left( \mathbf{q,x}\right) ,b_{N,M}\right]
&=&G_{M,M}\left( \mathbf{\bar{q}}\right) \frac{e^{-\frac{\bar{q}^{2}}{2}}}{%
\sqrt{N}}\sum_{p}\left[ G_{p,p}\left( \mathbf{\bar{Q}}^{\ast }\right)
-G_{p+N,p+N}\left( \mathbf{\bar{Q}}^{\ast }\right) \right] c_{p,M}^{\dagger
}c_{p+N,M} \\
&\simeq &e^{-\frac{\bar{q}^{2}}{2}}L_{M}\left( \frac{\bar{q}^{2}}{2}\right) %
\left[ L_{\nu -1}\left( \frac{\bar{Q}^{2}}{2}\right) -L_{\nu -1+N}\left(
\frac{\bar{Q}^{2}}{2}\right) \right] b_{N,M}\,,
\end{eqnarray*}
we finally conclude that $\hat{\rho}_{N}\left( \mathbf{q,x}\right) $ must
have the following bosonic representation:
\begin{equation}
\hat{\rho}_{N}\left( \mathbf{q,x}\right) =\left\langle G_{\mathbf{0}}\right|
\hat{\rho}_{N}\left( \mathbf{q,x}\right) \left| G_{\mathbf{0}}\right\rangle
+e^{-\frac{\bar{q}^{2}}{2}}\sum_{n,m}L_{m}\left( \frac{\bar{q}^{2}}{2}%
\right) \left[ L_{\nu -1}\left( \frac{\bar{Q}^{2}}{2}\right) -L_{\nu
-1+n}\left( \frac{\bar{Q}^{2}}{2}\right) \right] b_{n,m}^{\dagger }b_{n,m},
\label{roexb2}
\end{equation}
where
\begin{equation}
\left\langle G_{\mathbf{0}}\right| \hat{\rho}_{N}\left( \mathbf{q,x}\right)
\left| G_{\mathbf{0}}\right\rangle =\sum_{n=0}^{\nu -1}G_{n,n}\left( l%
\mathbf{q}^{\ast }-i\mathbf{x}^{\ast }/l\right) \sum_{m=0}^{N_{\Phi
}}G_{m,m}\left( l\mathbf{q}\right) =\frac{2\pi }{l^{2}}L_{\nu -1}^{1}\left(
\frac{x^{2}}{2l^{2}}\right) \delta \left( \mathbf{q}\right) \,.
\label{roexb3}
\end{equation}
In the last step we have used the relations (\ref{Fll}) and $%
\sum_{m=0}^{n}L_{m}^{\alpha }\left( x\right) =L_{n}^{\alpha +1}\left(
x\right) .$

The bosonic representation of $\rho \left( \mathbf{q}\right) $ is a
particular case of relation (\ref{rhoxi}) and is given by
\begin{equation}
\rho \left( \mathbf{q}\right) =\frac{e^{-\frac{\bar{q}^{2}}{4}}}{\sqrt{\bar{q%
}}}\sum_{n>0}\sqrt{n}\left[ G_{\nu -1,\nu -1+n}\left( \mathbf{\bar{q}}^{\ast
}\right) b_{n}\left( \bar{q}\right) +G_{\nu -1+n,\nu -1}\left( \mathbf{\bar{q%
}}^{\ast }\right) b_{n}^{\dagger }\left( \bar{q}\right) \right] +\hat{\rho}%
_{N}\left( \mathbf{q}\right) .  \label{rodir2}
\end{equation}
The last term, $\hat{\rho}_{N}\left( \mathbf{q}\right) =\hat{\rho}_{N}\left(
\mathbf{q,0}\right) $, will give a null contribution after its substitution
in (\ref{hdirro2}) if we neglect cubic terms in the boson operators. These
would describe the dissociation of an electron-hole pair \cite{Cheng}.

Substituting (\ref{rodir2}) in (\ref{hdirro2}), neglecting the cubic terms,
considering the system neutrality (that cancels the effect of the mean
electron density $\,\,\left\langle G_{\mathbf{0}}\right| \hat{\rho}%
_{N}\left( \mathbf{q}\right) \left| G_{\mathbf{0}}\right\rangle =\frac{2\pi
\nu }{l^{2}}\delta \left( \mathbf{q}\right) $ present in $\hat{\rho}%
_{N}\left( \mathbf{q}\right) =\hat{\rho}_{N}\left( \mathbf{q,0}\right) $,
eq.(\ref{roexb2})) and performing the angular integration over $arg\left(
\mathbf{q}\right) $ (that eliminates the non-diagonal quadratic terms, see
section \ref{Dia}), we finally obtain the direct term of (\ref{Htot}).

Now, following the same procedure and substituting the bosonic representation, (%
\ref{roexb1}) and (\ref{roexb2}), of $\rho \left( \mathbf{q,x}\right) $ in (%
\ref{HIEX}) we verify that, after the integration over $\arg \left( \mathbf{x%
}\right) $, the linear terms in the boson operators present null
coefficients. If we neglect cubic terms in the boson operators, we get $%
H_{I}^{Ex}$ written as:
\begin{equation*}
H_{I}^{Ex}=H_{Col}^{Ex}+H_{Self}^{Ex},
\end{equation*}
\begin{eqnarray}
H_{Col}^{Ex} &=&-\frac{1}{2\left( 2\pi \right) ^{2}}\int d^{2}q\frac{e^{-%
\frac{\bar{q}^{2}}{2}}}{\bar{q}}\int d^{2}xV\left( x\right) e^{-\frac{\bar{x}%
^{2}}{2}}e^{i(\mathbf{\bar{q}\cdot \bar{x}}^{\ast }-\mathbf{\bar{q}}^{\ast }%
\mathbf{\cdot \bar{x}})/2}  \notag \\
&&\times \left\{ \sum_{n>0}\sqrt{n}\left[ G_{\nu -1,\nu -1+n}\left( \mathbf{%
\bar{Q}}^{\ast }\right) b_{n}\left( \bar{q}\right) +G_{\nu -1+n,\nu
-1}\left( \mathbf{\bar{Q}}^{\ast }\right) b_{n}^{\dagger }\left( \bar{q}%
\right) \right] \right\} \times \left\{ h.c.\right\} ,   \label{HIEx2}
\end{eqnarray}
\begin{equation*}
H_{Self}^{Ex}=\frac{1}{2\pi l^{2}}\int d^{2}xV\left( x\right) e^{-\frac{\bar{%
x}^{2}}{2}}L_{\nu -1}^{1}\left( \frac{\bar{x}^{2}}{2}\right) \sum_{n>0}\left[
L_{\nu -1}\left( \frac{\bar{x}^{2}}{2}\right) -L_{\nu -1+n}\left( \frac{\bar{%
x}^{2}}{2}\right) \right] \int_{0}^{\infty }d\bar{q}b_{n}^{\dagger }\left(
\bar{q}\right) b_{n}\left( \bar{q}\right) .
\end{equation*}

Expanding eq.(\ref{HIEx2}) we will have different exchange effective
potentials, $v_{nn^{\prime }}^{Ex}(\mathbf{q)}$, associated with the
different combinations of operators $b_{n}$ and $b_{n^{\prime }}^{\dagger }$
obtained. For example, the one associated with the combination $b_{n}\left(
\bar{q}\right) b_{n^{\prime }}^{\dagger }\left( \bar{q}\right) $ is given
by:
\begin{equation}
v_{nn^{\prime }}^{Ex}(\mathbf{q)}=-\frac{\sqrt{nn^{\prime }}}{2\left( 2\pi
\right) ^{2}}\frac{e^{-\bar{q}/2}}{\bar{q}}\int d^{2}xV(x)e^{-\frac{\bar{x}%
^{2}}{2}}e^{i(\mathbf{\bar{q}\cdot \bar{x}}^{\ast }-\mathbf{\bar{q}\cdot
\bar{x}})/2}G_{\nu -1,\nu -1+n}(\mathbf{\bar{Q}}^{\ast })G_{\nu -1,\nu
-1+n^{\prime }}^{\ast }(\mathbf{\bar{Q}}^{\ast }).  \label{V1}
\end{equation}

The integration variable, $\mathbf{x}$, is different from the
variable of the
integrand, $\mathbf{\bar{Q}}^{\ast }\mathbf{=\bar{q}}^{\ast }-i\mathbf{\bar{x%
}}^{\ast }\mathbf{.}$ However, if we decompose the function $G(\mathbf{\bar{Q%
}})$ using the matrix product (\ref{Soma}), apply the Fourier
transform,
\begin{equation}
e^{-\bar{x}^{2}/2}G_{m,n}\left( -i\mathbf{\bar{x}}\right) G_{m^{\prime
},n^{\prime }}^{\ast }\left( -i\mathbf{\bar{x}}^{\ast }\right) =\frac{1}{%
2\pi }\int d^{2}\bar{k}e^{-i\mathbf{\bar{k}\cdot \bar{x}}}G_{n^{\prime
},n}\left( \mathbf{\bar{k}}^{\ast }\right) G_{m^{\prime },m}\left( \mathbf{%
\bar{k}}\right) e^{-\bar{k}^{2}/2},
\end{equation}
realize the integration over $\mathbf{\bar{x}}$ and, finally, contract the $%
G $ function using (\ref{Soma}) again, we get:
\begin{equation*}
v_{nn^{\prime }}^{Ex}(\mathbf{q)}=-\frac{\sqrt{nn^{\prime }}}{2\left( 2\pi
\right) ^{3}}\frac{1}{\bar{q}}\int d^{2}\bar{k}\tilde{V}(k)e^{-\bar{k}%
^{2}/2}e^{(\mathbf{\bar{q}\cdot \bar{k}}^{\ast }-\mathbf{\bar{k}\cdot \bar{q}%
}^{\ast })/2}G_{\nu -1+n^{\prime },\nu -1+n}(\mathbf{\bar{k}}^{\ast })G_{\nu
-1,\nu -1}(\mathbf{\bar{k}}^{\ast }).
\end{equation*}
\end{widetext}
The angular portion of this integral, $\mathbf{\bar{k}}=\bar{k}e^{i\theta _{%
\mathbf{k}}}$, is given by:
\begin{equation}
\int d\theta _{\mathbf{k}}e^{i\bar{k}(\cos \theta _{k}\bar{q}_{y}-\sin
\theta _{k}\bar{q}_{x})}e^{i\theta _{\mathbf{k}}\left( n-n^{\prime }\right)
}=2\pi \mathcal{J}_{n^{\prime }-n}\left( \bar{k}\bar{q}\right) e^{i\theta _{%
\mathbf{q}}\left( n^{\prime }-n\right) },  \label{An1}
\end{equation}
where we use eq.(3.937) from Ref.[39] and $\mathcal{J}_{n}$ is the
Bessel function of order $n$. Now, due the presence of the coefficient $%
e^{i\theta _{\mathbf{q}}\left( n^{\prime }-n\right) }$ in the expression
above, only the terms $n=n^{\prime }$ will survive after the angular
integration over $\arg \left( \mathbf{q}\right) =\theta _{\mathbf{q}}$ in (%
\ref{HIEx2}). Therefore, we will need only
\begin{eqnarray}
v_{n=n^{\prime }}^{Ex}(\mathbf{q)} &=&-\frac{n}{2\left( 2\pi \right) ^{2}}%
\frac{1}{\bar{q}}\int d\bar{k}\bar{k}\tilde{V}(k)e^{-\bar{k}^{2}/2}
\label{VBB} \\
&&\times L_{\nu -1+n}(\frac{\bar{k}^{2}}{2})L_{\nu -1}(\frac{\bar{k}^{2}}{2})%
\mathcal{J}_{0}\left( \bar{k}\bar{q}\right) .  \notag
\end{eqnarray}
Repeating the same procedure as above to the other coefficients, we verify
that the ones associated with the terms $b_{n}\left( q\right) b_{n^{\prime
}}\left( q\right) $ and $b_{n}^{\dagger }\left( q\right) b_{n^{\prime
}}^{\dagger }\left( q\right) $ will be proportional to $e^{\pm i\theta _{%
\mathbf{q}}\left( n^{\prime }+n\right) }$, instead of $e^{i\theta _{q}\left(
n^{\prime }-n\right) }$ as in (\ref{An1}). Therefore, once $n,n^{\prime }>0$%
, they will disappear after the angular integration over $\theta _{\mathbf{q}%
}$ in (\ref{HIEx2}). The coefficient associated with $b_{n}^{\dagger }\left(
q\right) b_{n^{\prime }}\left( q\right) $ will be the same as the one in (%
\ref{VBB}). In this way, remembering that $\tilde{V}(k)=2\pi
e^{2}/\varepsilon k$, we get the expressions (\ref{gex}) to (\ref{Vexn2}) .

\section{Calculation of the Functions $G_{m,\protect\theta%
}^{\gtrless}\left(t\right)$}

In this Appendix we present the calculation of the Green's functions $%
G_{m,\theta }^{\gtrless }\left( t\right) $, defined in (\ref{GnmT2}). The
method used is the common one in bosonization procedures. We suggest Ref.[16] for a review. We have
\begin{eqnarray*}
iG_{m,\theta }^{>}\left( t\right) &=&\left\langle \psi _{m}\left( \theta
,t\right) \psi _{m}^{\dagger }\left( 0,0\right) \right\rangle _{0} \\
&=&\lim_{\epsilon \rightarrow 0}\frac{1}{2\pi \epsilon }\left\langle
U_{m}\left( t\right) e^{-i\hat{N}_{m}\theta }e^{-i\Theta _{m}\left( \theta
,t\right) }e^{i\Theta _{m}\left( 0,0\right) }U_{m}^{\dagger }\right\rangle
_{0} \\
&=&\lim_{\epsilon \rightarrow 0}\frac{1}{2\pi \epsilon }e^{-i\left( \mu
_{m}^{+}t+\theta \right) }\left\langle e^{-i\Theta _{m}\left( \theta
,t\right) }e^{i\Theta _{m}\left( 0,0\right) }\right\rangle _{0},
\end{eqnarray*}
where we use (\ref{PotQ}). So, we just have to calculate the propagator
\begin{equation*}
K_{m,\theta }\left( t\right) =\lim_{\epsilon \rightarrow 0}\frac{1}{2\pi
\epsilon }\left\langle e^{-i\Theta _{m}\left( \theta ,t\right) }e^{i\Theta
_{m}\left( 0,0\right) }\right\rangle _{0}.
\end{equation*}

Using the identity $e^{A}e^{B}=e^{B}e^{A}e^{C}$, where $C=\left[
A,B\right]/2 $ and $\left[ C,A\right] =\left[ C,B\right] =0$, we
can write the above expression in a normal ordered form, or use
directly the relation
\begin{equation*}
\left\langle e^{\lambda _{1}\beta _{1}}e^{\lambda _{2}\beta
_{2}}\right\rangle _{0}=e^{\left\langle \lambda _{1}\beta _{1}\lambda
_{2}\beta _{2}+\frac{1}{2}\left( \lambda _{1}^{2}\beta _{1}^{2}+\lambda
_{2}^{2}\beta _{2}^{2}\right) \right\rangle _{0},}
\end{equation*}
valid for bosonic operators $\beta $ acting on eigenstates of a free boson
Hamiltonian \cite{VanD}, and we get:
\begin{equation*}
K_{m}\left( \theta ,t\right) =\lim_{\epsilon \rightarrow 0}\frac{1}{2\pi
\epsilon }e^{J_{m}\left( \theta ,t\right) },
\end{equation*}
where
\begin{equation*}
J_{m}\left( \theta ,t\right) =D_{m}\left( \theta ,t\right) -D_{m}\left(
0,0\right) ,
\end{equation*}
and
\begin{eqnarray*}
D_{m}\left( \theta ,t\right) &=&\left\langle \Theta _{m}\left( \theta
,t\right) \Theta _{m}\left( 0,0\right) \right\rangle \\
&=&\int_{0}^{\infty }d\bar{q}\Phi _{m}^{2}\left( \bar{q}\right)
\sum_{n=1}^{\infty }\frac{e^{-n\epsilon }}{n}e^{-in\theta }e^{-iE_{n}\left(
\bar{q}\right) t}.
\end{eqnarray*}
Here we have used the expression (\ref{FaseT}) of $\Theta _{m}\left( \theta
,t\right) $. Therefore,
\begin{eqnarray*}
iG_{m,\theta }^{>}\left( t\right) &=&\lim_{\epsilon \rightarrow 0}\frac{1}{%
2\pi \epsilon }e^{-i\left( \mu _{m}^{+}t+\theta \right) } \\
&&\times e^{\int_{0}^{\infty }d\bar{q}\Phi _{m}^{2}\left( \bar{q}\right)
\sum_{n=1}^{\infty }\frac{e^{-n\epsilon }}{n}\left[ e^{-in\theta
}e^{-iE_{n}\left( \bar{q}\right) t}-1\right] } \\
&=&\frac{1}{2\pi }e^{-i\left( \mu _{m}^{+}t+\theta \right)
}e^{\int_{0}^{\infty }d\bar{q}\Phi _{m}^{2}\left( \bar{q}\right)
\sum_{n=1}^{\infty }\frac{e^{-n\epsilon }}{n}e^{-in\theta }e^{-iE_{n}\left(
\bar{q}\right) t}},
\end{eqnarray*}
where we use $\ln \left( 1-x\right) =-\sum_{n=1}^{\infty }x^{n}/n$, with $%
x=e^{-\epsilon }$, in the last equation. The calculation of $G_{m,\theta
}^{<}\left( t\right) $ is analogous to the one presented above.

\section{Calculation of the Functions $G_{m,\protect\theta ,s}^{\gtrless
}\left( t\right) $}

Including the electron spin, the fermionic operator evolves in time as
\begin{eqnarray*}
\psi _{m,s}\left( \theta ,t\right) &=&\lim_{\epsilon \rightarrow 0}\frac{1}{%
\sqrt{2\pi \epsilon }}U_{m,s}\left( t\right) e^{-i\hat{N}_{m,s}\theta
}e^{-i\Theta _{m,s}\left( \theta ,t\right) }, \\
\Theta _{m,s}\left( \theta ,t\right) &=&i\sum_{n=1}^{\infty }\frac{%
e^{-n\epsilon /2}}{\sqrt{n}}\int_{0}^{\infty }d\bar{q}\Phi _{m}\left( \bar{q}%
\right) \\
&&\times \left[ e^{-in\theta }b_{n,s}\left( \bar{q},t\right) -e^{in\theta
}b_{n,s}^{\dagger }\left( \bar{q},t\right) \right] .
\end{eqnarray*}

Using (\ref{OBCS1}) and (\ref{OBCS2}) we can express the operators $%
b_{n,s}\left( \bar{q}\right) $ in terms of the charge and spin bosonic
operators, $b_{n}\left( \bar{q}\right) _{\rho }$ and $b_{n}\left( \bar{q}%
\right) _{\sigma }$. Then we will have
\begin{eqnarray*}
\Theta _{m,s}\left( \theta ,t\right) &=&-\frac{i}{\sqrt{2}}%
\sum_{n=1}^{\infty }\frac{e^{-n\epsilon /2}}{\sqrt{n}}\int_{0}^{\infty }d%
\bar{q}\Phi _{m}\left( \bar{q}\right) \\
&&\times\left\{ e^{in\theta }b_{n}^{\dagger }\left( \bar{q},t\right) _{\rho
}-e^{-in\theta }b_{n}\left( \bar{q},t\right) _{\rho }\right. \\
&&\left. +s\left[ e^{in\theta }b_{n}^{\dagger }\left( \bar{q},t\right)
_{\sigma }-e^{-in\theta }b_{n}\left( \bar{q},t\right) _{\sigma }\right]
\right\} .
\end{eqnarray*}
Due the separation of the total Hamiltonian $H_{T}$ in the charge and spin
modes (see (\ref{HTTS})) we will have the trivial evolution, $b_{n}\left(
\bar{q},t\right) _{i}=e^{-iE_{n}^{\left( i\right) }\left( \bar{q}\right)
t}b_{n}\left( \bar{q}\right) _{i}$, with $E_{n}^{\left( i\right) }\left(
\bar{q}\right) $ given by (\ref{HTS}).

Now, the retarded Green's function associated with the inclusion of a
particle in the quantum state $\left| n,m,s\right\rangle $ is given by:
\begin{eqnarray}
iG_{n,m,s}^{Ret}\left( t\right) &=&\theta \left( t\right) \left\langle
c_{n,m,s}\left( t\right) c_{n,m,s}^{\dagger }\left( 0\right) \right.  \notag
\\
&&\left. +c_{n,m,s}^{\dagger }\left( 0\right) c_{n,m,s}\left( t\right)
\right\rangle  \label{Grets} \\
&=&i\theta \left( t\right) \int_{0}^{2\pi }d\theta e^{in\theta }\left[
G_{m,s,\theta }^{>}\left( t\right) +G_{m,s,\theta }^{<}\left( t\right) %
\right] ,  \notag
\end{eqnarray}
where
\begin{eqnarray}
iG_{m,s,\theta }^{>}\left( t\right) &=&\left\langle \psi _{m,s}\left( \theta
,t\right) \psi _{m,s}^{\dagger }\left( 0,0\right) \right\rangle ,  \notag \\
iG_{m,s,\theta }^{<}\left( t\right) &=&\left\langle \psi _{m,s}^{\dagger
}\left( 0,0\right) \psi _{m,s}\left( \theta ,t\right) \right\rangle .
\label{Gmom}
\end{eqnarray}

Now, due the commutativity between the operators $b_{n}^{(\dagger )}\left(
\bar{q},t\right) _{\rho }$ and $b_{n}^{(\dagger )}\left( \bar{q},t\right)
_{\sigma }$ and the separation of the total Hamiltonian in the form (\ref
{HTTS}), we have the characteristic factorization \cite{voit,Schon} of the
operators $G_{m,s,\theta }^{\gtrless }\left( t\right) $ into charge and spin
propagators. For example:
\begin{eqnarray*}
iG_{m,s,\theta }^{>}\left( t\right) &=&e^{-i\mu _{s}^{+}t}e^{-i\theta
}K_{m,\rho ,\theta }\left( t\right) K_{m,\sigma ,\theta }\left( t\right) , \\
K_{m,i,\theta }\left( t\right) &=&\frac{1}{\sqrt{2\pi }}e^{\frac{1}{2}%
\sum_{n=1}^{\infty }\frac{e^{-n\epsilon /2}}{n}e^{-in\theta
}\int_{0}^{\infty }d\bar{q}\Phi _{m}^{2}\left( \bar{q}\right)
e^{-iE_{n}^{\left( i\right) }\left( \bar{q}\right) t}},
\end{eqnarray*}
where the factor $1/2$ in the exponent comes from the factor $1/\sqrt{2}$ in
the definitions (\ref{OBCS1}) and (\ref{OBCS2}) of $b_{n}\left( \bar{q}%
,t\right) _{i}.$ Calculating $iG_{m,s,\theta }^{<}\left( t\right) $ in a
analogous way and substituting in (\ref{Grets}) we get (\ref{Gretnms}), with
\begin{equation*}
M_{m,\theta }\left( t\right) =K_{m,\rho ,\theta }\left( t\right) K_{m,\sigma
,\theta }\left( t\right) \,.
\end{equation*}


\begin{thebibliography}{99}
\bibitem{Obs}  When we say $\nu =1$ we means, in the Kallin and Halperin
notation, $\nu _{\downarrow }=\nu _{\uparrow }=1$, once we are always
considering the two spin states (Landau sublevels) initially equally
occupied.
\end{thebibliography}
\end{document}